\documentclass[twocolumn,letter,prc,preprintnumbers,superscriptaddress,showpacs,amsmath,amssymb,floatfix]{revtex4}
\usepackage{graphicx}
\usepackage{dcolumn}
\usepackage{colordvi}
\usepackage{longtable}
\begin{document}
\title{Cross sections and beam asymmetries for $\vec{e} p \to e n \pi^+$ in the nucleon resonance region for $1.7 \leq Q^2 \leq 4.5~(\rm{GeV})^2$}

\date{\today}

\newcommand*{\KYUNGPOOK}{Kyungpook National University, Daegu 702-701, Republic of Korea}
\affiliation{\KYUNGPOOK}
\newcommand*{\JLAB}{Thomas Jefferson National Accelerator Facility, Newport News, Virginia 23606}
\affiliation{\JLAB}
\newcommand*{\ANL}{Argonne National Laboratory}
\affiliation{\ANL}
\newcommand*{\ASU}{Arizona State University, Tempe, Arizona 85287-1504}
\affiliation{\ASU}
\newcommand*{\UCLA}{University of California at Los Angeles, Los Angeles, California  90095-1547}
\affiliation{\UCLA}
\newcommand*{\CSU}{California State University, Dominguez Hills, Carson, CA 90747}
\affiliation{\CSU}
\newcommand*{\CMU}{Carnegie Mellon University, Pittsburgh, Pennsylvania 15213}
\affiliation{\CMU}
\newcommand*{\CUA}{Catholic University of America, Washington, D.C. 20064}
\affiliation{\CUA}
\newcommand*{\SACLAY}{CEA-Saclay, Service de Physique Nucl\'eaire, 91191 Gif-sur-Yvette, France}
\affiliation{\SACLAY}
\newcommand*{\CNU}{Christopher Newport University, Newport News, Virginia 23606}
\affiliation{\CNU}
\newcommand*{\UCONN}{University of Connecticut, Storrs, Connecticut 06269}
\affiliation{\UCONN}
\newcommand*{\ECOSSE}{Edinburgh-Glasgow Admin.}
\affiliation{\ECOSSE}
\newcommand*{\ECOSSEE}{Edinburgh University, Edinburgh EH9 3JZ, United Kingdom}
\affiliation{\ECOSSEE}
\newcommand*{\EMMY}{Emmy-Noether Foundation, Germany}
\affiliation{\EMMY}
\newcommand*{\FU}{Fairfield University, Fairfield CT 06824}
\affiliation{\FU}
\newcommand*{\FIU}{Florida International University, Miami, Florida 33199}
\affiliation{\FIU}
\newcommand*{\FSU}{Florida State University, Tallahassee, Florida 32306}
\affiliation{\FSU}
\newcommand*{\GWU}{The George Washington University, Washington, DC 20052}
\affiliation{\GWU}
\newcommand*{\ECOSSEG}{University of Glasgow, Glasgow G12 8QQ, United Kingdom}
\affiliation{\ECOSSEG}
\newcommand*{\ISU}{Idaho State University, Pocatello, Idaho 83209}
\affiliation{\ISU}
\newcommand*{\INFNFR}{INFN, Laboratori Nazionali di Frascati, 00044 Frascati, Italy}
\affiliation{\INFNFR}
\newcommand*{\INFNGE}{INFN, Sezione di Genova, 16146 Genova, Italy}
\affiliation{\INFNGE}
\newcommand*{\ORSAY}{Institut de Physique Nucleaire ORSAY, Orsay, France}
\affiliation{\ORSAY}
\newcommand*{\ITEP}{Institute of Theoretical and Experimental Physics, Moscow, 117259, Russia}
\affiliation{\ITEP}
\newcommand*{\JMU}{James Madison University, Harrisonburg, Virginia 22807}
\affiliation{\JMU}
\newcommand*{\MIT}{Massachusetts Institute of Technology, Cambridge, Massachusetts  02139-4307}
\affiliation{\MIT}
\newcommand*{\UMASS}{University of Massachusetts, Amherst, Massachusetts  01003}
\affiliation{\UMASS}
\newcommand*{\MOSCOW}{Moscow State University, General Nuclear Physics Institute, 119899 Moscow, Russia}
\affiliation{\MOSCOW}
\newcommand*{\UNH}{University of New Hampshire, Durham, New Hampshire 03824-3568}
\affiliation{\UNH}
\newcommand*{\NSU}{Norfolk State University, Norfolk, Virginia 23504}
\affiliation{\NSU}
\newcommand*{\OHIOU}{Ohio University, Athens, Ohio  45701}
\affiliation{\OHIOU}
\newcommand*{\ODU}{Old Dominion University, Norfolk, Virginia 23529}
\affiliation{\ODU}
\newcommand*{\PITT}{University of Pittsburgh, Pittsburgh, Pennsylvania 15260}
\affiliation{\PITT}
\newcommand*{\RPI}{Rensselaer Polytechnic Institute, Troy, New York 12180-3590}
\affiliation{\RPI}
\newcommand*{\RICE}{Rice University, Houston, Texas 77005-1892}
\affiliation{\RICE}
\newcommand*{\URICH}{University of Richmond, Richmond, Virginia 23173}
\affiliation{\URICH}
\newcommand*{\SCAROLINA}{University of South Carolina, Columbia, South Carolina 29208}
\affiliation{\SCAROLINA}
\newcommand*{\TRIUMF}{TRIUMF, 4004, Wesbrook Mall, Vancouver, BC, V6T 2A3, Canada}
\affiliation{\TRIUMF}
\newcommand*{\UNIONC}{Union College, Schenectady, NY 12308}
\affiliation{\UNIONC}
\newcommand*{\VT}{Virginia Polytechnic Institute and State University, Blacksburg, Virginia   24061-0435}
\affiliation{\VT}
\newcommand*{\VIRGINIA}{University of Virginia, Charlottesville, Virginia 22901}
\affiliation{\VIRGINIA}
\newcommand*{\WM}{College of William and Mary, Williamsburg, Virginia 23187-8795}
\affiliation{\WM}
\newcommand*{\YEREVAN}{Yerevan Physics Institute, 375036 Yerevan, Armenia}
\affiliation{\YEREVAN}
\newcommand*{\NOWUNH}{University of New Hampshire, Durham, New Hampshire 03824-3568}
\newcommand*{\NOWMOSCOW}{Moscow State University, General Nuclear Physics Institute, 119899 Moscow, Russia}
\newcommand*{\NOWSCAROLINA}{University of South Carolina, Columbia, South Carolina 29208}
\newcommand*{\NOWUMASS}{University of Massachusetts, Amherst, Massachusetts  01003}
\newcommand*{\NOWMIT}{Massachusetts Institute of Technology, Cambridge, Massachusetts  02139-4307}
\newcommand*{\NOWECOSSEE}{Edinburgh University, Edinburgh EH9 3JZ, United Kingdom}
\newcommand*{\NOWECOSSEG}{University of Glasgow, Glasgow G12 8QQ, United Kingdom}

\author {K.~Park} 
\altaffiliation[Current address:]{\NOWSCAROLINA}
\affiliation{\KYUNGPOOK}
\affiliation{\JLAB}
\author{V.D.~Burkert}
     \affiliation{\JLAB}
\author{W.~Kim}
     \affiliation{\KYUNGPOOK}
\author{I.G.~Aznauryan}
     \affiliation{\YEREVAN}
     \affiliation{\JLAB}
\author{R.~Minehart}
     \affiliation{\VIRGINIA}
\author{L.C.~Smith}
     \affiliation{\VIRGINIA}
\author{K.~Joo}
     \affiliation{\UCONN}
     \affiliation{\VIRGINIA}
\author{L.~Elouadrhiri}
     \affiliation{\CNU}
     \affiliation{\JLAB}
\author {G.~Adams} 
\affiliation{\RPI}
\author {M.J.~Amaryan} 
\affiliation{\ODU}
% ...
 
\collaboration{The CLAS Collaboration}
     \noaffiliation

\begin{abstract}
The exclusive electroproduction process $\vec{e}p \to e^\prime n \pi^+$ was measured in the range of the photon virtuality $Q^2 = 1.7 - 4.5~\rm{GeV^2}$, and the invariant mass range for the $n\pi^+$ system of $W = 1.15 - 1.7~\rm{GeV}$ using the CEBAF Large Acceptance Spectrometer. For the first time, these kinematics are probed in exclusive $\pi^+$ production from protons with nearly full coverage in the azimuthal and polar angles of the $n\pi^+$ center-of-mass system. The $n\pi^+$ channel has particular sensitivity to the isospin $\frac{1}{2}$ excited nucleon states, and together with the $p\pi^0$ final state will serve to determine the transition form factors of a large number of resonances. The largest discrepancy between these results and present modes was seen in the $\sigma_{LT'}$ structure function. In this experiment, 31,295 cross section and 4,184 asymmetry data points were measured. Because of the large volume of data, only a reduced set of structure functions and Legendre polynomial moments can be presented that are obtained in model-independent fits to the differential cross sections. 
\end{abstract}

\maketitle

\section{Introduction}
The study of the excited states of the nucleon is an important step in the development of a fundamental understanding of the strong interaction~\cite{isgur_2000}. While the existing data of the low-lying resonances are consistent with the well-studied $SU(6) \otimes O(3)$ constituent quark model classification, many open questions remain. On the fundamental level there exists only a very limited understanding of the relationship between Quantum Chromo-Dynamics(QCD), the field theory of the strong interaction, and the constituent quark models or alternative hadron models. Experimentally, we still do not have sufficiently complete data that can be used to uncover unambiguously the structure of the nucleon and its excited states. For a recent overview of results available before 2004 see Ref. ~\cite{burkert_lee_2004}. Precise data to study the transition from the nucleon ground state to the $\Delta(1232)$, in the electroproduction of $\pi^0$ with large-range angular coverage and in a wide range of photon virtualities have become available in recent years~\cite{Joo2002,Joo2003,Sparveris2003,Biselli2003,Joo2004,Kelly2005,Stave06,Ungaro06,Sparveris06}. Electromagnetic multipoles have been extracted from these measurements covering a large range in photon virtuality $0 \leq Q^2 \leq 6~\rm{GeV^2}$. These results have proved crucial in advancing the development of lattice QCD methods to study $\gamma^* N \Delta$ transition form factors~\cite{Alexandrou04,Alexandrou05}.

The $\Delta(1232)$ state is a relatively isolated isospin $\frac{3}{2}$ resonance and is quite accessible in $\pi^0$ electroproduction from proton targets. In the mass region above the $\Delta(1232)$, there is a cluster of three nucleon resonances, the $N(1440)$, $N(1520)$  and $N(1535)$ in the mass range around 1.5 $\rm{GeV}$, and at least nine $N^*$ and $\Delta^*$ states in a mass range from 1.62 to 1.72$~\rm{GeV}$, many of them with large branching ratios into the $N\pi$ hadronic final state. Single pion electroproduction is highly sensitive to many of these states. In order to disentangle the different states through their isospin and spin-parity assignments, more detailed experimental information is needed than is available from the $p\pi^0$ final state alone. In particular, most of the states with masses up to 1.7 $\rm{GeV}$ have isospin $\frac{1}{2}$, and couple more strongly to the $n\pi^+$ final state than to $p\pi^0$. A detailed mapping of this channel is crucial for a successful analysis of the mass range above the $\Delta(1232)$. Such an analysis requires complete information on the center-of-mass angle distribution to separate the contributing partial waves. The first measurement of exclusive $\pi^+$ electroproduction from protons at low $Q^2$ in the resonance region and with complete angular coverage has become available only recently~\cite{HEgiyan2006}. Previous measurements~\cite{Breuker} were very limited in angle coverage and statistical accuracy. Moreover, measurements of polarization observables are very important. Their sensitivity to interferences of resonant and non-resonant amplitudes can enhance the contributions of smaller resonances. For the exclusive $n\pi^+$ final state, beam polarization asymmetries have only been measured in the lower mass and $Q^2$ region ~\cite{Joo2005}, and double polarization observables are available only in limited kinematics ~\cite{Devita02}. 

The symmetric constituent quark model(CQM) allows one to make predictions for the systematics of the excited $N^*$ and $\Delta^*$ spectrum, as well as for the internal structure of these states. While the resonance spectrum up to a mass of 1.7~$\rm{GeV}$ is reasonably well explored, the internal structure of most states above the $\Delta(1232)$ has only been studied very crudely. For example, the lowest nucleon-like state is the $N(1440)$ with $J^P =\frac{1}{2}^+$. Model predictions for this state disagree widely on its transition form factors, and precise experimental information is currently available only from single pion photoproduction~\cite{Arndt2002,pdg2004}, and in the range $Q^2 < 0.65~\rm{GeV^2}$ from recent single pion~\cite{HEgiyan2006} and double pion electroproduction~\cite{Ripani2003}. The analyses of these data~\cite{Aznauryan2005_1,Aznauryan2005_2} made use of differential cross sections as well as of polarized electron beam asymmetries. The latter were found highly sensitive to the amplitudes of the very broad $N(1440)$ state through interference of the resonant and nonresonant amplitudes. They revealed transition form factors that show a very strong $Q^2$ dependence for the transverse(magnetic) amplitude, and a large coupling to longitudinal photons. Such a behavior is not understood within non-relativistic CQMs~\cite{Li_Close,Pfeil} or the hybrid model ~\cite{Li_Burkert_Li}, and indicates possible large contributions from vector mesons~\cite{Cano} or relativistic effects~\cite{Capstick_Keister}. To further explore this behavior, measurements at higher $Q^2$ are necessary, where models make distinctly different predictions. 

The transition to the $N(1520)$ state with $J^P = \frac{3}{2}^-$ is predicted within the CQM~\cite{Close_Gilman,Capstick_Keister,Giannini} to rapidly change the helicity structure of the $\gamma NN^*$  vertex from the total helicity $\lambda_{\gamma N} = \frac{3}{2}$ dominance at the real photon point to $\lambda_{\gamma N}= \frac{1}{2}$ dominance at short distances(i.e. high $Q^2$).  Quark models predict a similar behavior for the $N(1680)$ $J^P = \frac{5}{2}^+$ state. Earlier analyses of older data found indications for such a behavior ~\cite{burkert_lee_2004,Foster}, but a precise mapping over a large $Q^2$ range has not been accomplished. Apart from the $\Delta(1232)$, the $N(1535)$ is the only state for which the transverse transition form factor has been measured in a large range of $Q^2$~\cite{Thompson2001,Armstrong1999,burkert_lee_2004,Thompson2007}. This state has a large branching ratio to both the $p\eta$ and the $N\pi$ channels. Measurement of this state in the $n\pi$ channel is important to obtain information on the longitudinal photocoupling amplitude that is difficult to access in the $p\eta$ channel. Moreover, it will allow us to test for meson cloud effects in the resonance transition, which may be different for the two channels.

\section{Kinematics}
\begin{figure}[!htb]
\begin{center}
	\includegraphics[angle=0,width=0.35\textwidth]{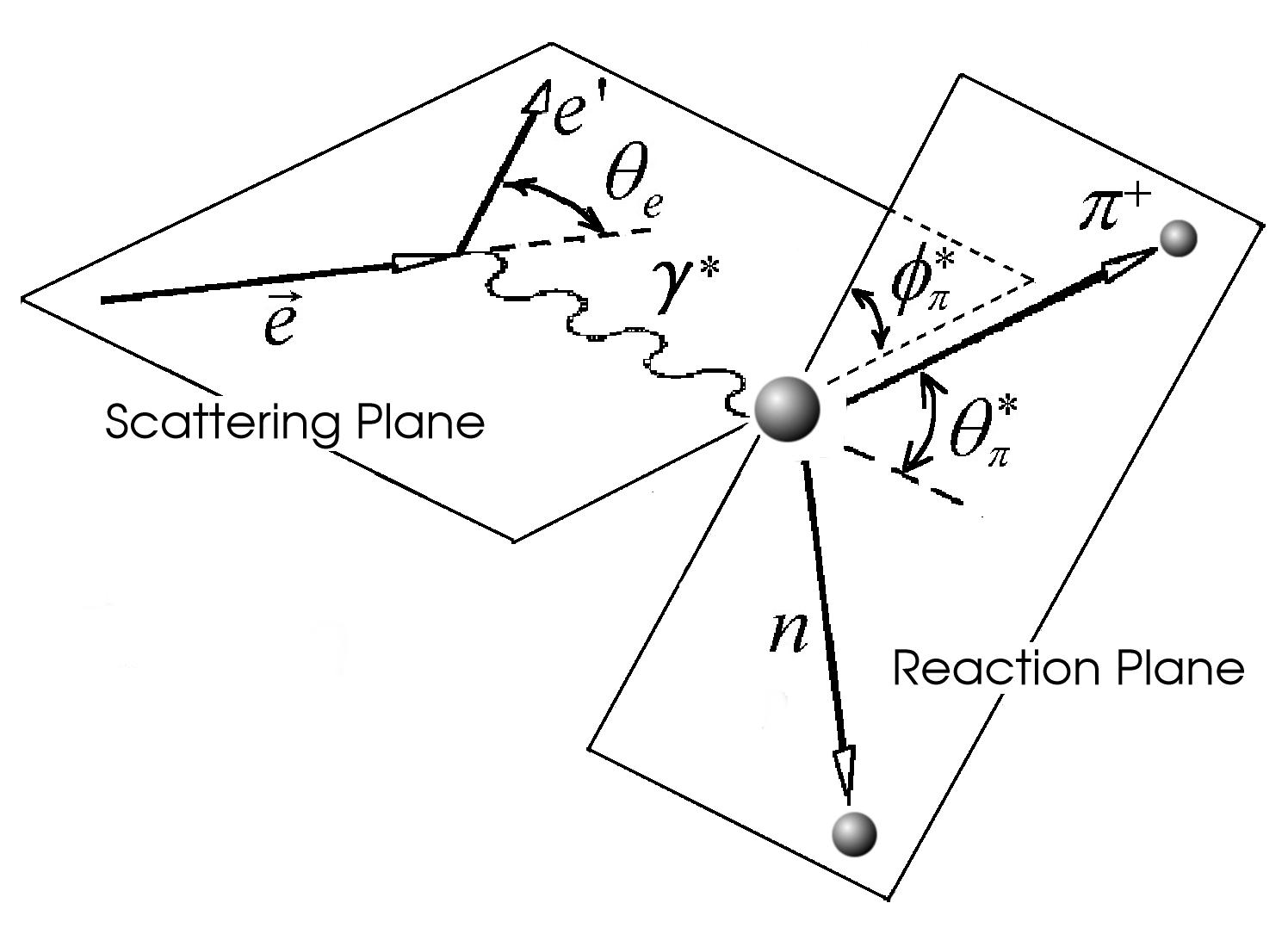}
        \caption{
          Kinematics of single $\pi^+$ electroproduction.}
          \label{fig:pion_kinematics}
\end{center}
\end{figure}

We report on measurements of differential cross sections and polarized electron beam asymmetries with the CEBAF Large Acceptance Spectrometer(CLAS) at Jefferson Lab using a polarized continuous electron beam of 5.754 $\rm{GeV}$ energy incident upon a liquid-hydrogen target. The kinematics of single pion electroproduction is displayed in Fig.~\ref{fig:pion_kinematics}.
 In the one-photon exchange approximation, the electron kinematics is described by two Lorentz invariants: $Q^2$, characterizing the virtuality of the exchanged photon, and $\nu$, the transferred energy. 
\begin{eqnarray} 
Q^2 \equiv -(k_i-k_f)^2 = 4E_iE_f\sin^2{\theta_e \over 2}\\
\nu \equiv \frac{p_i\cdot p_{\gamma}}{M_p} = E_i - E_f ~,
\end{eqnarray}
where $k_i$ and $k_f$ are the initial and final four momenta of the electron, and $p_{\gamma}$ and $p_i$ are the virtual photon and target four momenta. $E_i$ and $E_f$ are the initial and final electron energies in the laboratory frame, $\theta_e$ is the electron scattering angle, and $M_p$ is the proton mass. Another related quantity is the invariant mass of the hadronic final state $W$ that can be expressed as: 
\begin{eqnarray}
W^2 \equiv (p_{\gamma} + p_i)^2 = M_p^2 + 2M_p\nu - Q^2~.
\end{eqnarray}        
In this measurement the scattered electron and the outgoing $\pi^+$ are detected, while the final state neutron is unobserved. Since the 4-momentum of the incident electron and of the target proton are known, the 4-momentum of the missing system $X$  in the final state can be reconstructed and its mass determined as:  
\begin{eqnarray}
M_X^2 \equiv ((k_i + p_i) - (k_f + q_{\pi}))^2~.
\label{eqn:miss_mass}
\end{eqnarray}
 where, $q_{\pi}$ is the pion 4-vector. For single $\pi^+$ production, the constraint on the missing mass is $M_X = M_n$. The outgoing $\pi^+$ is defined by two angles in the center-of-mass frame, the polar angle $\theta^*_{\pi}$ and the azimuthal angle $\phi^*_{\pi}$. The latter is the angle between the electron scattering plane and the hadronic production plane. It is defined such that the scattered electron trajectory lies in the $\phi^*_{\pi} = 0$ half plane with the z-axis pointing along the virtual photon 3-momentum vector.  The kinematics is completely defined by five variables $(Q^2, W, \theta^*_{\pi}, \phi^*_{\pi}, \phi_e)$. The $\phi_e$ is the electron azimuthal angle. In the absence of a transverse polarization of the beam or the target nucleon, the cross section does not depend on $\phi_e$, and can be written as~\cite{burkert_lee_2004}:
\begin{eqnarray}
\frac{\partial^5\sigma}{\partial E_f \partial\Omega_e \partial\Omega^*_{\pi}} &=& \Gamma_v \cdot \frac{d^2\sigma}{d\Omega^*_{\pi}}
\end{eqnarray}
where, 
\begin{eqnarray} 
\Gamma_v & =& \frac{\alpha}{2\pi^2Q^2} \frac{(W^2-M_p^2)E_f}{2M_pE_e} \frac{1}{1-\epsilon}\\
\epsilon &=& (1 + 2(1 + \frac{\nu^2}{Q^2})\tan^2{\theta_e\over 2})^{-1}      \\
\frac{d^2\sigma } {d\Omega_{\pi}^*} &=& \sigma_T + \epsilon \sigma_L + \epsilon \sigma_{TT}\cos{2\phi^*_{\pi}} + \sqrt{2\epsilon(1+\epsilon)} \sigma_{LT}\cos{\phi^*_{\pi}} \nonumber\\
 &+& h \sqrt{2\epsilon(1-\epsilon)} \sigma_{LT'} \sin{\phi^*_{\pi}} \label{eqn:cs_form}
\end{eqnarray}
The parameter $\epsilon$ represents the virtual photon polarization, $\Gamma_v$ is the virtual photon flux, $h$ is the electron helicity and $\frac{d\sigma}{d\Omega^*_{\pi}}$ is the differential pion photoabsorption cross section.
\section{Reaction Models \label{sect:models}}

 Beginning in the late 1990's, model approaches have been developed that aim at accurately reproducing the experimental data. In section~\ref{sec:results} we compare some of the results with calculations based on model descriptions such as the Dubna-Mainz-Taipei~(DMT) model~\cite{dmt00}, several versions of the MAID model~\cite{maid2000}, and the Sato-Lee model~(SL)~\cite{satolee01}. In addition, a unitary isobar model(UIM) was developed by the Yerevan-JLab group~\cite{IGAzn,Aznauryan2005_1} that contains many features of MAID, but incorporates different energy-dependences of the background amplitudes. This approach allows us to fit experimental cross sections and polarization asymmetries to extract resonance transition form factors. We briefly summarize the main features of these models. They are discussed in more detail in Ref. ~\cite{burkert_lee_2004}. 

\begin{figure}[htb]
\begin{center}
	\includegraphics[angle=0,width=0.35\textwidth]{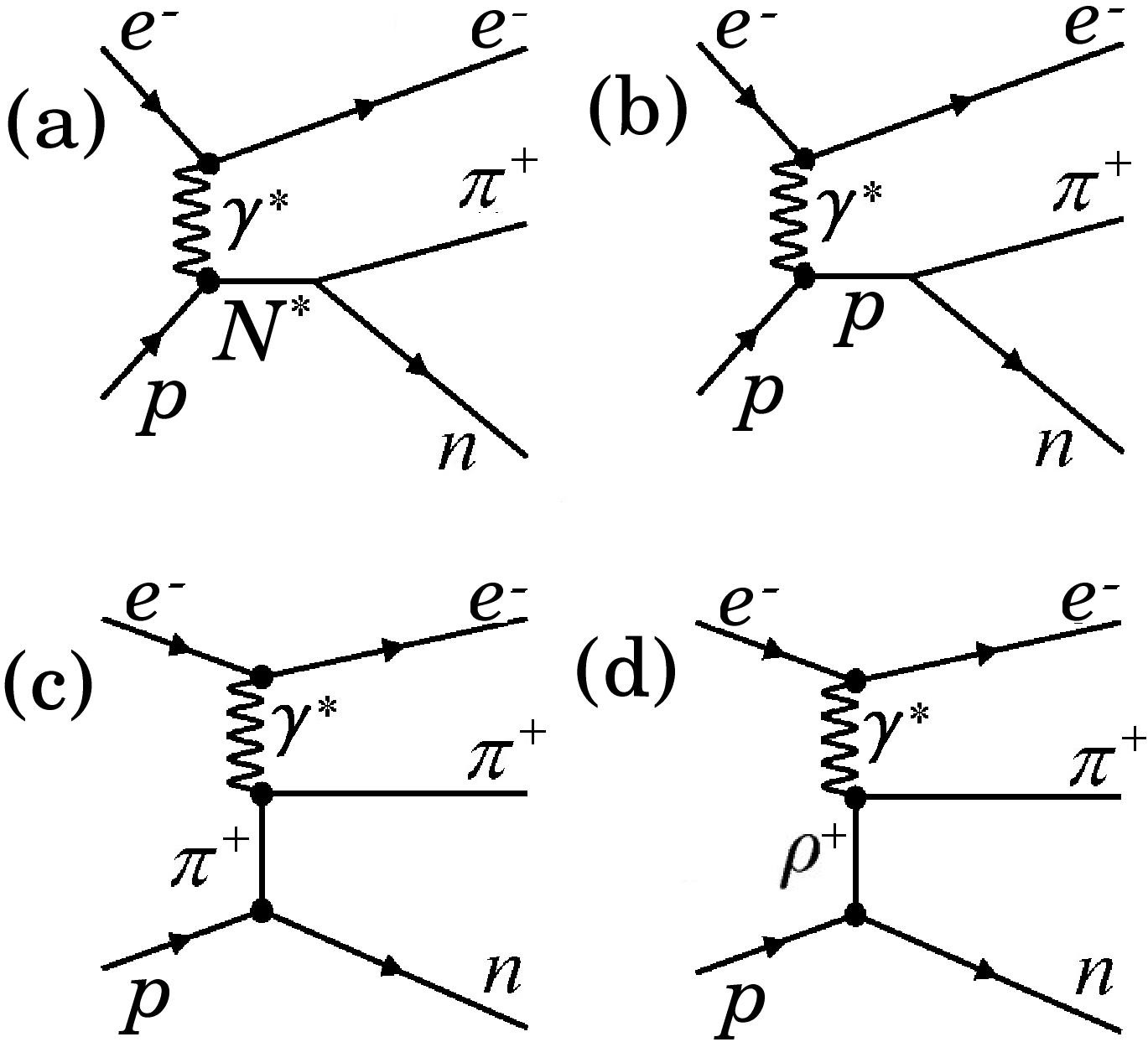}
        \caption{
          Tree level description of single $\pi^+$ electroproduction. (a) $s-$channel resonance production (b) $s-$channel nucleon exchange (c) $t-$channel pion exchange (d) $t-$channel $\rho$ meson exchange}
          \label{fig:Born_terms}
\end{center}
\end{figure}

MAID and related models are based on an isobar description of the single pion production process. They incorporate non-resonant amplitudes described by tree level Born terms, and also include $\rho$ and $\omega$ t-channel processes that are relevant mostly in the region of higher resonances. Figure~\ref{fig:Born_terms} shows the diagrams contributing to the reaction $e p \to e n \pi^+$ at low and intermediate energies. The vertex functions for the virtual photon coupling to hadrons are parameterized according to their respective on-shell form factors for which there is prior experimental information. Resonances are parameterized by a phenomenological description using a relativistic Breit-Wigner form with an energy-dependent width. The total amplitude for single pion production is unitarized in a K-matrix formulation. Only single channels are included, and multi-channel effects such as $\gamma N \to (\rho N, \pi\Delta) \to \pi N$, which could be important in the second and third resonance regions, are neglected. From an experimental viewpoint, the attractive feature of these descriptions is their flexibility that allows adjusting parameters, such as electromagnetic transition form factors and hadronic couplings, as new experimental information becomes available. However, all of these descriptions lack significant predictive power, and a comparison with new data will tell us more about how well electromagnetic and hadronic couplings have been parameterized, rather than about the intrinsic structure of the nucleon.         

Dynamical models, such as the SL and DMT models start from a consistent Hamiltonian formulation. In these models the non-resonant interaction modifies the resonant amplitude. The SL model provides the most consistent description of the interaction, but is currently limited to the region of the $\Delta(1232)$ resonance, while in the DMT model the resonance amplitudes are parameterized according to a specific Breit-Wigner form that simplifies the inclusion of higher resonances. The s-channel resonance parameterization in the DMT model is similar to what is used in the isobar descriptions such as the MAID and UIM approaches. A different approach is used in the Ohio model ~\cite{ohio_model}, which starts from a  Salpeter equation for the pion-nucleon system. The photon is subsequently attached to describe the photo-pion reaction.   In this approximation, retardation effects are neglected, and the pion, nucleon, and resonance exchanges appear instantaneously. In pion electroproduction this approach leads to an unphysical singularity at finite $Q^2$, which can be avoided in some ad hoc approximations. The model has so far  only been used to compare with polarization beam asymmetries at relatively low photon virtualities~\cite{ohio_2005}.   

Once the transition form factors have been extracted from the data, their interpretation in terms of the intrinsic structure of the nucleon must then involve comparisons with nucleon structure models, such as the many versions of CQM's, and with Lattice QCD calculations.  

\section{Experimental Setup} 
\begin{figure}[!htb]
\begin{center}
	\includegraphics[angle=0,width=0.50\textwidth]{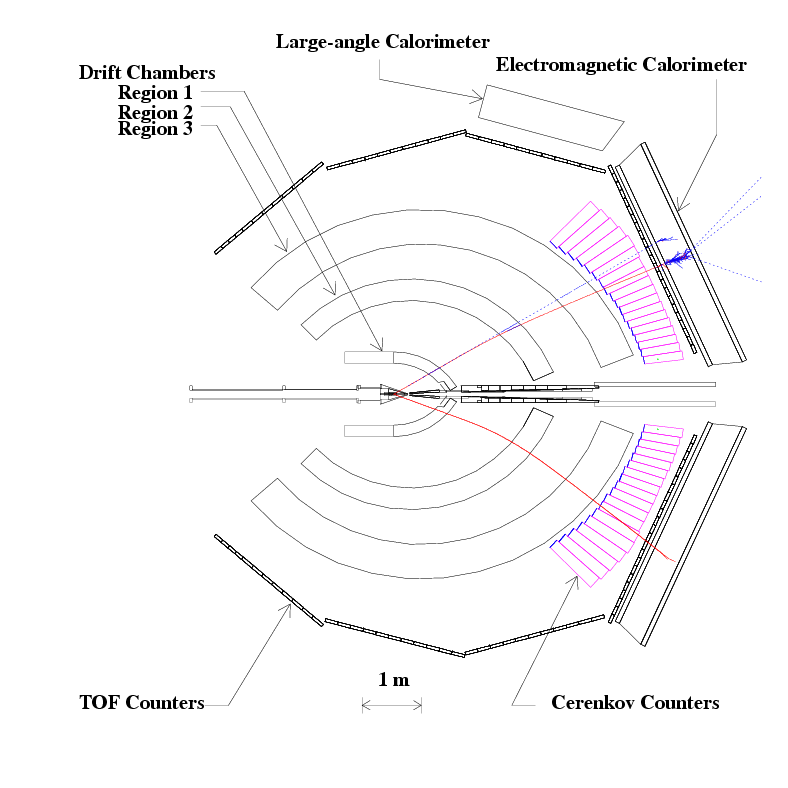}
	\includegraphics[angle=0,width=0.46\textwidth]{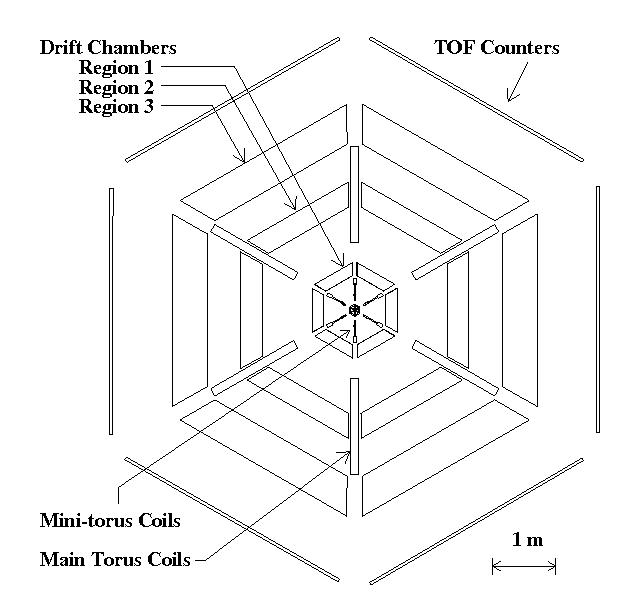}
        \caption{
	  (Color online) Schematics of the CLAS detector system. The top panel shows a 
	cut through sectors 1 and 4 along the beam line. The beam enters from the left into CLAS. A GEANT-simulated event is shown with an electron bending towards the beam line and a positive particle in the opposite sector bending away from the beam. The bottom panel shows a cut perpendicular to the beam line through the center of CLAS.   
          \label{fig:clas}
        }
\end{center}
\end{figure}

The measurement was carried out with the CEBAF Large Acceptance Spectrometer(CLAS)~\cite{clas}. A schematic view of CLAS is shown in Fig.~\ref{fig:clas}. CLAS utilizes a magnetic field distribution generated by six flat superconducting coils, arranged symmetrically in azimuth. The coils generate an approximate toroidal field distribution around the beam axis. The six identical sectors of the magnet are independently instrumented with 34 layers of drift cells for particle tracking, plastic scintillation counters for time-of-flight(TOF) measurements, gas threshold $\check{\rm C}$erenkov counters(CC) for electron and pion separation and triggering purposes, and scintillator-lead sampling calorimeters(EC) for photon and neutron detection and triggering. To aid in electron/pion separation, the EC is segmented into an inner part facing the target, and an outer part away from the target. CLAS covers on average 80\% of the full 4$\pi$ solid angle for the detection of charged particles. Azimuthal angle acceptance is maximum at large polar angles and decreases at forward angles. Polar angle coverage ranges from about 8$^{\circ}$ to 140$^{\circ}$ for the detection of $\pi^+$. Electrons are detected in the CC and EC covering polar angles from 15$^{\circ}$ to 55$^{\circ}$, this range being somewhat dependent on the momentum of the scattered electron. The target is surrounded by a small toroidal magnet with normal-conducting coils. This magnet is used to shield the drift chambers closest to the target from the intense low-energy electron background resulting from Moller electron scattering processes. 
In the current experiment, only two charged particles need to be detected, the scattered electron and the produced $\pi^+$, while the full final state is reconstructed using four-momentum conservation constraints. The continuous electron beam provided by CEBAF is well suited for measurements involving two or more final state particles in coincidence, leading to very small accidental coincidence contributions of  $< 10^{-3}$ for the instantaneous luminosity of $10^{34}$cm$^{-2}$sec$^{-1}$ used in this measurement.   

The measurement was performed from October 2001 to January 2002. A polarized electron beam of 8 nA current and an energy of 5.754 $\rm{GeV}$ was directed onto a 5~cm long liquid-hydrogen target. The longitudinally average beam polarization was 72.7\% and was routinely measured during the experiment using a M\/oller electron polarimeter. The beam helicity was switched at a rate of 30~sec$^{-1}$ in a pseudo-random fashion, and the charge for each helicity state was integrated in a totally absorbing Faraday cup(FC). Empty target runs were performed to measure contributions from the target cell windows. The target was located 4~cm upstream of the nominal CLAS center. The torus magnet was set at 90\% of its maximum field. Events were triggered on a single electron candidate defined as a concidence of the total energy deposited in one sector of the EC and a signal in the CC in the same sector. A minimum energy of 640 $\rm{MeV}$ deposited in one EC sector was required in the trigger. All events were first written to a RAID disk array, and later transferred to the tape silo of the Jefferson Lab computer center. Raw data were subjected to the calibration and reconstruction procedure that are part of the standard CLAS data analysis chain. The reaction studied in this paper contributed only a fraction to the total event sample, and a more stringent event selection(``skimming'') was applied to select events with one electron candidate and only one positively charged track. These events were subject to further selection criteria described in the following sections. 

\section{Data Analysis}
\subsection{Event selection}
\subsubsection{Electron identification}
Selection of electron candidates in CLAS at the level 1 trigger is achieved by requiring that the energy deposited in the EC and the CC hit be in the same sector. Such an open trigger does not provide a stringent electron selection at the relatively high beam energy, and additional selection criteria must be applied in the offline event analysis. First, we require that the EC and CC hits are geometrically matched with a negatively charged track reconstructed in the drift chambers(DC). Secondly, we employ the direct correlation between the energy deposited in the calorimeter and the momentum obtained in the track reconstruction in the magnetic field. About 30\% of the total energy deposited in the EC is directly measured in the active scintillator material. This detectable portion of the EM shower is referred to as the ``sampling fraction''($\alpha$). The remaining 70\% of the energy is deposited mostly in the lead sheets that are interleaved between the scintillator sheets as showering material. A GEANT~\cite{geant} based Monte Carlo simulation package(GSIM) was used to determine the EC response as a function of electron energy.  The sampling fraction is nearly energy-independent, and for this experiment $\alpha \equiv E_{vis}/E_{tot} = 0.291$ where, $E_{vis}$ is the visible deposited energy in the scintillator material, $E_{tot}$ is the total deposited energy in the scintillator material of EC. The value of $\alpha$ can vary somewhat with the energy calibration of the calorimeter, but was kept constant during the entire run period. Lower values of $\alpha$ are observed in cases where electrons hit the calorimeter near the edges, and a fraction of the shower energy leaks out of the calorimeter volume. In order to eliminate such edge effects, fiducial regions were defined for the calorimeter that assure full energy response as long as the electrons hit the calorimeter inside the fiducial regions.  

\begin{figure}[!thb]
\begin{center}
  \includegraphics[angle=0,width=0.48\textwidth]{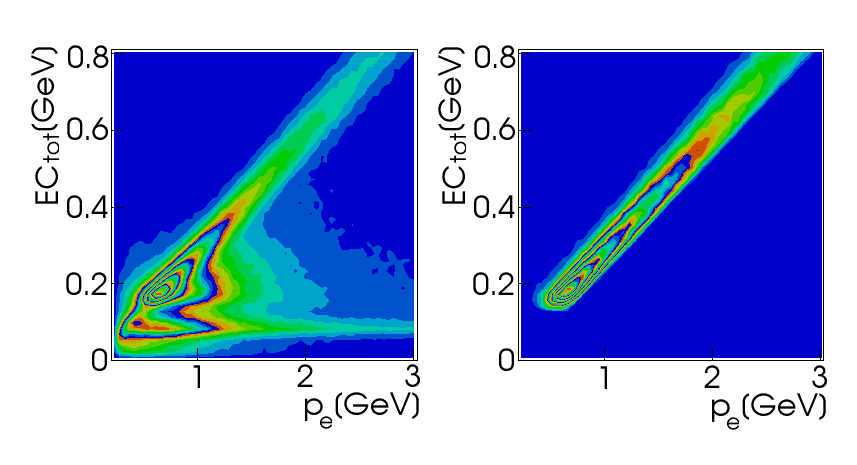}
\caption{(Color online) Energy in EC vs momentum for electron candidates before EC energy cuts(left), and after all cuts(right). 
\label{fig:ec_momentum}}
\end{center}
\end{figure}

In contrast to electrons, charged pions do not create EM showers and deposit energy largely though ionization, resulting in minimum energy deposited in the calorimeter. Minimum ionizing pions are easily eliminated by simple minimum energy cuts.  Pions that undergo hadronic interactions also deposit only a fraction of their full energy in the calorimeter volume, with more energy lost in the outer parts of the EC, while showering electrons deposit most of their energy in the inner part of the calorimeter. Cuts were applied to the sampling ratio as well as to the minimum energy deposited in the total EC and in the inner part($EC_{inner}$). Figure~\ref{fig:ec_momentum} shows the total energy deposited in the EC scintillators versus the electron momentum before and after all cuts are applied to the sampling ratio and the total EC energy. Pions were rejected with the following energy cuts: $EC_{inner} >$ 50 MeV and $EC_{total} >$ 140 MeV.   In addition, events were eliminated if the number of photoelectrons recorded in the CC did not exceed 2.5 for electron candidates. Such tracks were more likely associated with negatively charged pions than with electrons. Using a Poisson distribution for the number of photoelectrons, corrections were applied for the small losses of electron events that occurred due to this cut. These corrections were done separately for all bins in $\theta$ and $\phi$ angles to take into account the variation of the average number of photoelectrons with kinematics.

\begin{figure}[htb]
\begin{center}
	\includegraphics[angle=0,width=0.47\textwidth]{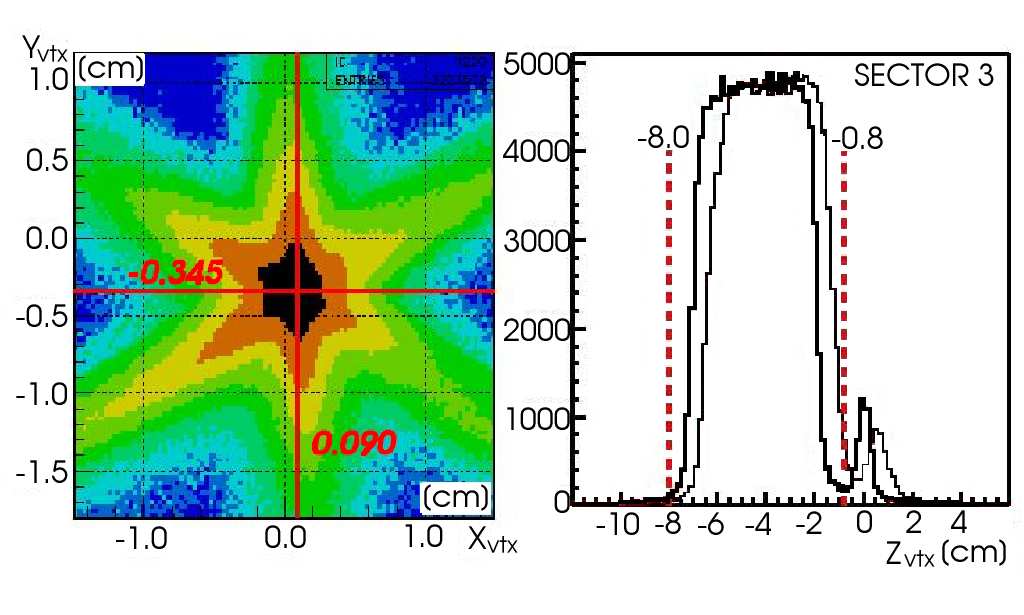}
        \caption{(Color online) Left: Reconstructed x and y target positions, showing an offset of x=0.90 mm and y=-3.45 mm, respectively. The right panel shows the z-vertex before(thin solid line) and after(bold solid line) the beam offsets in the x and y target positions have been corrected. 
          \label{fig:xyzvertex}
        }
\end{center}
\end{figure}

The electron beam was centered on the hydrogen production target cell which, as can be seen in Fig.~\ref{fig:xyzvertex}, was located vertically about -3.5 mm relative to the CLAS center, and horizontally displaced by about 0.9 mm. The beam offset caused an azimuthal dependence of the reconstructed $z$-vertex($z_{vtx}$). After the beam offset was corrected, the azimuthal dependence of $z_{vtx}$ was eliminated. Events were selected in the range $-80 < z_{vtx} < -8$~mm to eliminate contributions from the exit window of the scattering chamber, which is located 2~cm downstream of the target cell. 
\subsubsection{Pion identification}        

\begin{figure}[!htb]
\begin{center}
        \includegraphics[angle=0,width=0.35\textwidth]{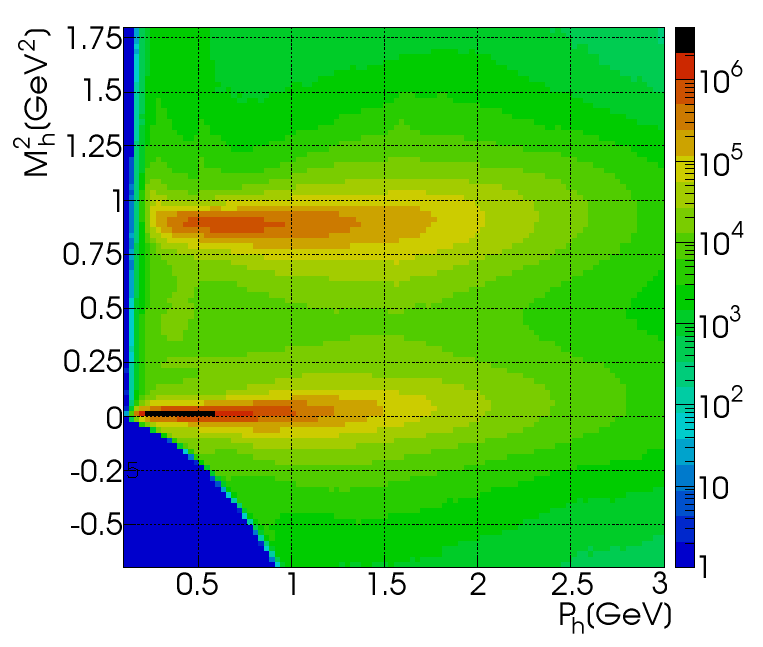}
        \caption{
         \protect
          (Color online)  Reconstructed hadron mass(squared) vs hadron momentum. The pion and proton mass bands are clearly visible.
         \label{fig:pion_id_beta_mean}
        }
\end{center}
\end{figure}

 Charged pions are identified by combining the particle velocity $\beta = v/c$, which is obtained from the difference of the vertex start time and the time-of-flight measurement in the TOF counters with the particle momentum from tracking through the magnetic field using the CLAS drift chamber system.
Precise timing calibrations are obtained by relating the electron timing to the highly stabilized radio frequency of the CEBAF accelerator. In order to isolate pions from protons, a 3$\sigma$ cut on $\beta_h$ vs. $p_h$ is applied. Using the reconstructed particle momentum and the timing information from the time-of-flight counters, the mass of the particle was determined, and is displayed versus the particle momentum in the Fig.~\ref{fig:pion_id_beta_mean}.  
After pions were selected, the start time of the event at the vertex was determined using the reconstructed pathlength of the pion track and the timing in the TOF scintillator paddles. A time resolution of $\delta T_e \approx 150~\rm{ps}$ was achieved. The vertex start time is needed to determine the velocity of the charged hadrons in the event.

\begin{figure}[htb]
\begin{center}
        \includegraphics[angle=0,width=0.38\textwidth]{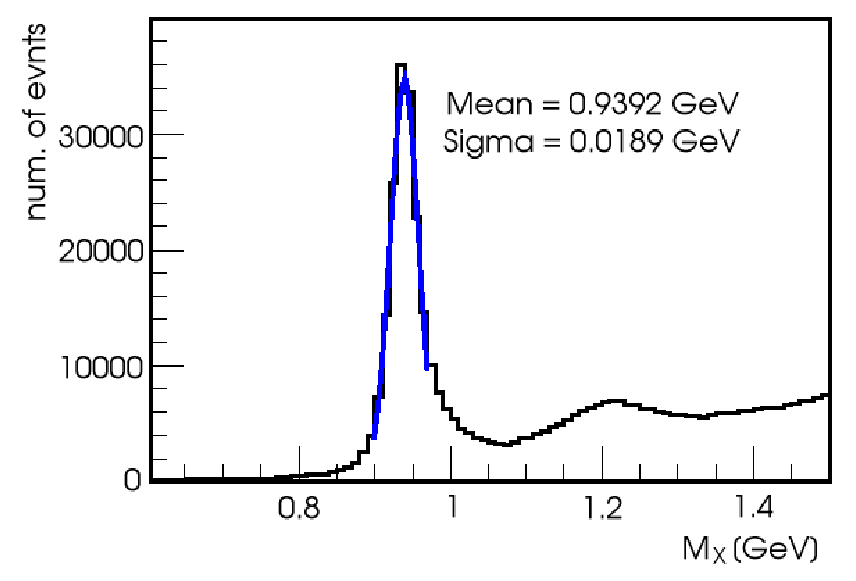}
        \caption{
         Missing mass $M_X$ distribution for $e p \to e \pi^+ X$ events.
         \label{fig:mmx_resol}
        }
\end{center}
\end{figure}

\subsection{Channel identification}
 The final state neutron is not directly observed in this experiment. However, the four-momentum vectors of all other particles are known and four-momentum conservation and charge conservation allow the determination of the charge and the mass of the unmeasured part of the final state. The exclusive process $e p \to e \pi^+ n$ is then identified by a peak in the missing mass distribution.  An example of the event distribution versus missing mass $M_X$ is shown in Figure~\ref{fig:mmx_resol}. The narrow peak at the nucleon mass indicates the exclusive process we aim to measure. We note that there is negligible accidental background visible under the peak, which would show up as a broad distribution below the neutron mass. The tail at the higher mass side of the neutron peak is due to radiative processes. The broad enhancement near 1.2~$\rm{GeV}$ is due to the process $e p \to e \pi^+ \Delta^0(1232)$. In order to select the exclusive process with the missing neutron in the final state, the neutron peak in each kinematical bin is fitted with a Gaussian distribution, and a 3$\sigma$ cut is applied to separate the $n \pi^+$ final state from double pion production $\pi^+ (\pi N)$. This cut also eliminates some events that are part of the radiative tail for single pion production. These losses have to be corrected for when extracting the unradiated cross section. These corrections are discussed in section~\ref{sec:radiative_corrections}. 

\subsection{Kinematic corrections}
Evidence for the need of kinematical corrections is seen in the dependence of the invariant mass of the elastic peak on the azimuthal angle. This effect is most prominent at forward polar angles where the torus coils come close to each other, and is largely due to small misalignments of the torus coils resulting in a slightly asymmetric magnetic field distribution. To compensate for the small magnetic field distortions, corrections were made to the reconstructed particle momentum vector. As a first step we use the kinematically constrained elastic $ep \to ep$ process to correct for possible distortions in the reconstructed scattering angle.  Using the known electron beam energy, the elastic ep scattering kinematics is completely determined by the two angles. The proton angle is well measured at large scattering angles where the tracking system is well aligned, and we assume it to be accurately known, while scattered electrons are detected at small angle where the alignment of the tracking chambers is less well known, and small position shifts can result in significant shifts in reconstructed angles. Given these conditions, the electron scattering angle can then be predicted and compared with the measured angle.  The corrections turn out to be less than 1 mrad for most of the phase space, however close to the torus coils, corrections can be up to 5~mrad. We attribute this significant effect to the distortions of the magnetic field close to the torus coils. Electron momentum corrections are derived from the difference between the predicted and measured momenta, using the corrected polar angles for elastically scattered electrons.  The size of these corrections decreases to less than 0.5\% with increasing scattering angle, but can be up 1.5\% close to the torus coils. Corrections to the polar angle of the $\pi^+$ are applied using the angle corrections previously determined for electrons. The $\pi^+$ momentum is corrected by matching the observed missing mass $M_X$ to the neutron mass in the process $e p \to e \pi^+ X$. The exclusive process $e p \to e \pi^+ n$ is determined with a neutron mass resolution of $\sigma_n \approx 18~\rm{MeV}$. 

The kinematic corrections have been tested using other exclusive processes with a neutral particle in the final state, e.g. $e p \to e p \pi^{\circ}$, $e p \to e p \eta$, and $e p \to e p \omega$. In all cases, the mass of the undetected particles is reconstructed with better than 2 $\rm{MeV}$ accuracy. We take this as evidence that the kinematics of the measured particles are well determined after all corrections are applied.  

\subsection{Fiducial volumes}

The $e p \to e \pi^+ n$ reaction has been simulated in the entire phase space allowed by the incident beam energy and the CLAS acceptance. However, the CLAS acceptance is a complicated function of the kinematical variables, and there are areas, e.g. the mechanical support structure of the $\check{\rm C}$erenkov counter mirrors and areas close to the CLAS torus coils, that are difficult to model with GSIM. To avoid the complication of edge effects, fiducial volumes with nominally full acceptance for particle detection were defined. These functions depend on azimuthal and polar angles, momentum, and charge, and are different for electrons and pions.
\begin{figure}[!htb]
\begin{center}
        \includegraphics[angle=0,width=0.47\textwidth]{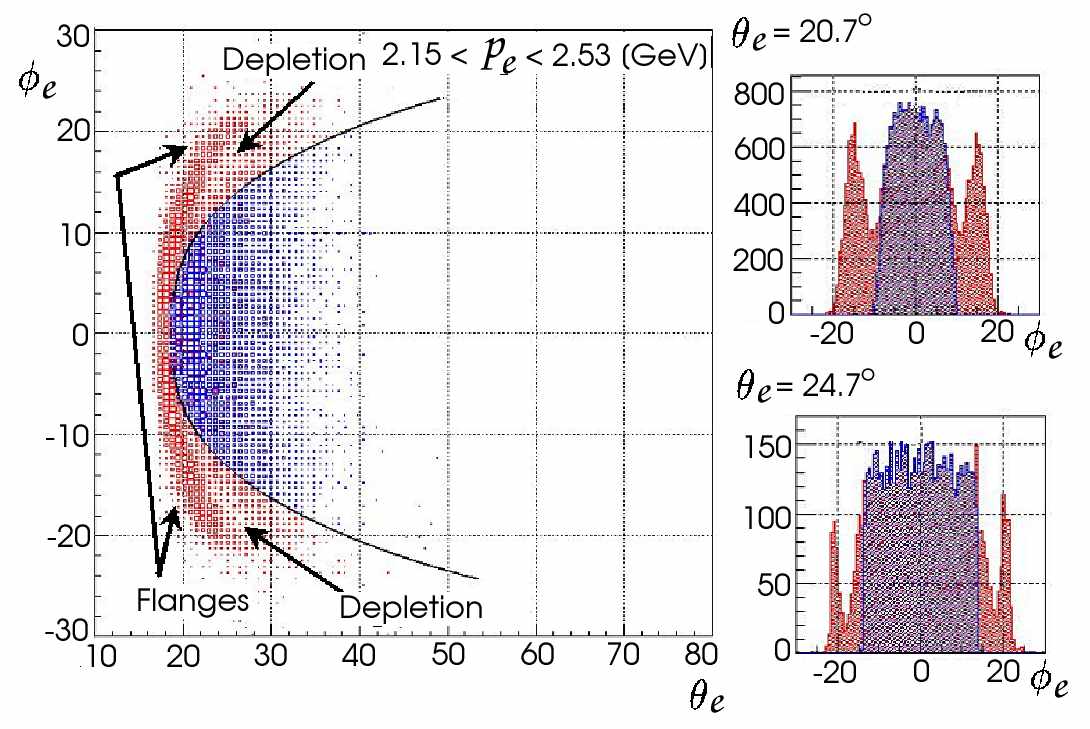}
        \caption{\protect
          (Color online) Electron fiducial cut at electron momentum range : $2.15 < p_e < 2.53~\rm{GeV}$ for sector 1. 
         The histograms on the right show the $\phi_e$ distributions at two values of $\theta_e$. The highlighted area 
	in the center indicates the selected fiducial range.
          \label{fig:electron_fcut}
        }
\end{center}
\end{figure}

\subsubsection{Electron fiducial volumes}
Geometrical fiducial cuts were defined to select forward regions of the detector that could be reliably simulated by the GSIM program. The $\check{\rm C}$erenkov counter efficiency has a complicated dependence on $\theta_e$ and $\phi_e$ near the acceptance edges. Fiducial volumes were defined to isolate the regions with uniform efficiency distributions. Due to the effects of the magnetic field, the angular fiducial volume also depends on the momentum of the scattered electron. The electron($\theta_e$, $\phi_e$) distributions are shown in Fig.~\ref{fig:electron_fcut} without and with the fiducial cuts applied. At forward angles a rapidly varying response of the $\check{\rm C}$erenkov counters can be seen that is due to non-uniform light collection. Applying the fiducial volume cut eliminates these regions from further analysis. 
The solid curve in Fig.~\ref{fig:electron_fcut} shows the boundary of the fiducial cut for the central momentum in that bin. Only events inside the black curve(\Blue{blue area}) are used in the analysis. In addition, a set of $\theta_e$ versus $p_e$ cuts was used to eliminate areas with reduced efficiency due to malfunctioning time-of-flight counter photomultipliers or drift chamber segments with broken wires. The CLAS detector also contains regions with no acceptance or with low efficiency. These regions were removed as well. Holes in the acceptance are mainly due to the torus coils, and in the forward region, due to the vacuum beam pipe and lead shielding surrounding the beam pipe.  

\subsubsection{Pion fiducial volumes}

The fiducial volumes for the produced $\pi^+$ are significantly different from the electron fiducial volumes. Since pion detection requires only charge particle tracking in the drift chamber system and time-of-flight measurements in the plastic scintillators, pions were detected in a much larger polar angle range from about 8$^{\circ}$ to 140$^{\circ}$.   Pion acceptance at low angles is increased by the fact that pions are bend away from the beamline.

\subsection{Kinematical binning}
\begin{figure}[!thb]
\begin{center}
        \includegraphics[angle=0,width=0.39\textwidth]{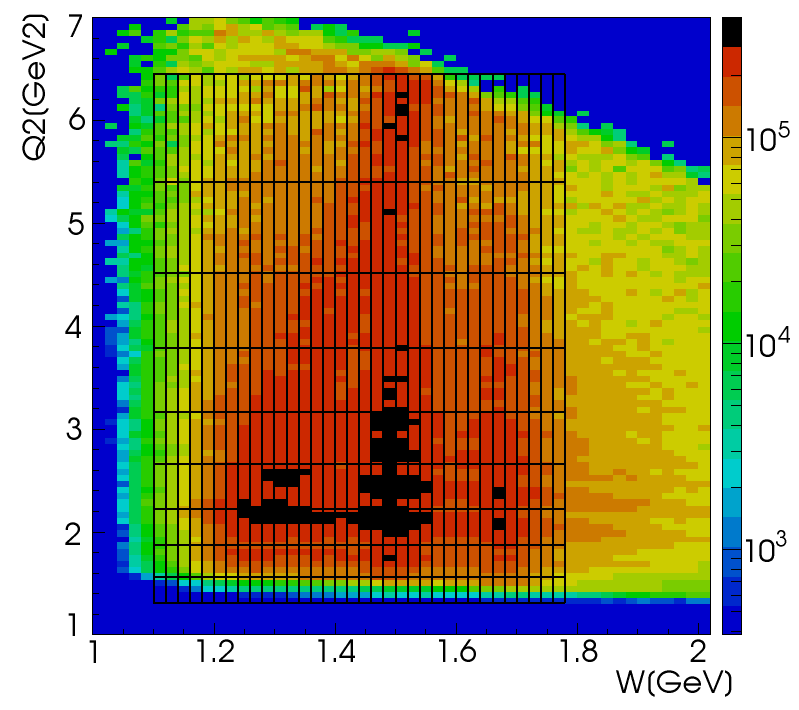}
        \includegraphics[angle=0,width=0.39\textwidth]{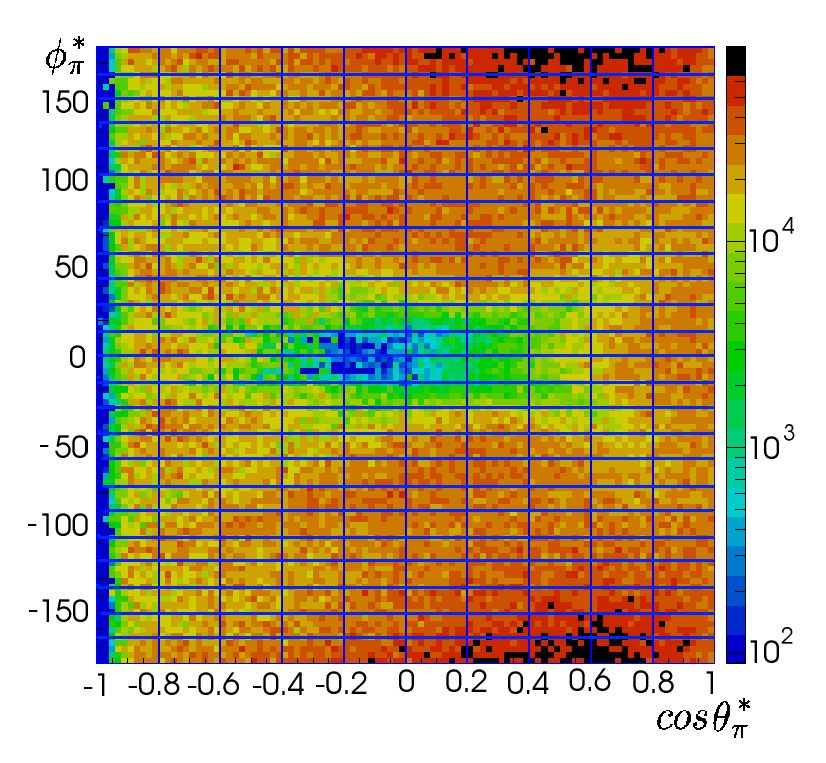}
        \caption{
          (Color online) Kinematic coverage in $Q^2\;[\rm{GeV^2}]$ versus $W\;[\rm{GeV}]$(top) and in ${\phi^*_{\pi}}\;\rm{^o}$ versus $\cos{\theta^*_{\pi}}$(bottom). 
	The solid lines show the bins used in the data analysis. 
          \label{fig:kinebin}
        }
\end{center}
\end{figure}

The CLAS detector covers a very large kinematic range in the four variables $W,~Q^2,~\cos{\theta^*_{\pi}}$ and $~\phi^*_{\pi}$. For further analysis, the data binning was matched to the underlying physics to be extracted. The study of nucleon excitations requires the analysis of the azimuthal $\phi^*_{\pi}$ dependence of the differential cross section to determine structure functions in the differential cross section, and the analysis of the polar angle dependence to identify the partial wave contributions at a given invariant mass of the hadronic final state. The binning in the hadronic mass $W$ must accommodate variations in the cross section, taking into account the width of resonances and their threshold behavior. On the other hand, the $Q^2$-dependence is expected to be smooth. Table~\ref{tab:kine_range} shows the binning in these variables. The $Q^2$ binning varies as $\Delta{Q^2} = 0.2 \cdot Q^2$ to partly compensate for the rapid drop in cross section with increasing $Q^2$, while the binning in the other quantities is fixed. The total number of cross section bins is 45,360. Figure~\ref{fig:kinebin} shows coverage in the hadronic center-of-mass angles, and the binning used for the extraction of the differential cross sections. As can be seen, the measurement covers nearly the entire range in $\phi^*_{\pi}$ and $\cos\theta^*_{\pi}$, with the exception of a region near $\phi^*_{\pi} = 0^{\circ}$ and $\cos\theta^*_{\pi} = -0.2$, where the acceptance is significantly reduced. These regions are eliminated from the analysis by requiring a minimum acceptance for each bin. 

\begin{table}[!htb]
\caption{Kinematical binning}
\begin{center}
\begin{tabular}{|c|c|c|c|}
\hline
Variable & Num. Bin & Range & Bin Size \\
\hline
\hline
$W$  & 27  & $1.15 - 1.7\;\rm{GeV}$ & $20\;\rm{MeV}$ \\
\hline
$Q^2$ & 7 & $1.1 - 5.0\;\rm{GeV^2}$ & variable \\
\hline
$\cos{\theta^*_{\pi}}$ & 10 & $-1.0 - 1.0$ & 0.2\\  
\hline
${\phi^*_{\pi}}$ & 24 & $-180. - 180^o$ & $15^o$ \\
\hline
\end{tabular}
\label{tab:kine_range}
\end{center}
\end{table}

\section{Simulations}
An essential part of the data analysis is the accurate modeling of the acceptance and event reconstruction efficiency for the process $ep \to e\pi^+n$ in the entire kinematic region accessible with CLAS.  The MAID00 physics model~\cite{maid2000} was used as an event generator to populate the covered phase space as closely as possible to the measured distributions.  Nearly 200M $ep \to e \pi^+ n$ events were generated covering the measured kinematics. A GSIM Post Processor(GPP) was used to adjust the detector response such that the simulated resolutions were compatible in their widths with the measured distributions. This allowed us to apply the same selection criteria for the simulated events as for the data, and gave an accurate estimate of acceptances and reconstruction efficiencies. The GPP was also used to account for missing channels in the drift chambers, and malfunctioning photomultipliers and electronics channels in the various detectors.  As previously discussed, cuts were applied to limit the reconstructed events to the fiducial volumes.
 
\subsection{Acceptance corrections}  
The CLAS detector has a large acceptance, however there are important non-uniformities and inefficiencies in some areas that need to be carefully taken into account when relating the experimentally measured yields to the differential cross sections. The complexity of the geometrical acceptance convoluted with the reconstruction efficiency that depends on all kinematical variables, prohibits an analytical parameterization of the detector response. Instead, for each of the 58,800 kinematic bins in $Q^2,~W,~\cos\theta^*_{\pi}$ and $~\phi^*_{\pi}$, a single number was determined which represents the combined acceptance and efficiency for this particular bin. In addition to the acceptance corrections, the data need to be corrected for radiative effects, which were included in the simulations. The number of acceptance-corrected events in each bin is given by:
\begin{eqnarray}
N_{corr} = N_{exp}/Acc~~~~ Acc = \frac{REC_{RAD}}{THR_{UNRAD}} ~,
\end{eqnarray}
where  $THR_{UNRAD}$ is the number of generated un-radiative events, $REC_{RAD}$ is the number of radiative events reconstructed in the simulation, $N_{exp}$ is the number of experimentally observed events, $Acc$ is the acceptance factor, and  $N_{corr}$ is the number of acceptance-corrected and de-radiated events. The latter includes all effects related to the detector resolution, e.g. event migration from the bin the event was generated to another bin where it was reconstructed. 

In some regions, for example close to the torus coils, the acceptance may change rapidly with the azimuthal angle $\phi^*_{\pi}$, and may even be zero in part of the bin. To avoid inaccuracies of the acceptance calculations due to the binning effects, we placed minimum acceptance cut 2.5\% all bins. This cut affected mostly the region near $\phi^*_{\pi} = 0^{\circ}$. The average acceptance is around $6 \sim 7\%$.

\subsection{Radiative corrections \label{sec:radiative_corrections}}

The inclusive radiative corrections developed by Mo and Tsai~\cite{mo_tsai} cannot be applied to exclusive pion electroproduction without additional assumptions. In this analysis we have used the approach developed by Afanasiev {\em et al.}~\cite{afanasev} for exclusive electroproduction of pseudoscalar mesons. 
This approach uses a model cross section as input, and performs an exact calculation without relying on the usual ``peaking approximation'' or the separate treatment of ``soft'' and ``hard'' photon radiation. 

\begin{figure}[!thb]
\begin{center}
	\includegraphics[angle=0,width=0.23\textwidth]{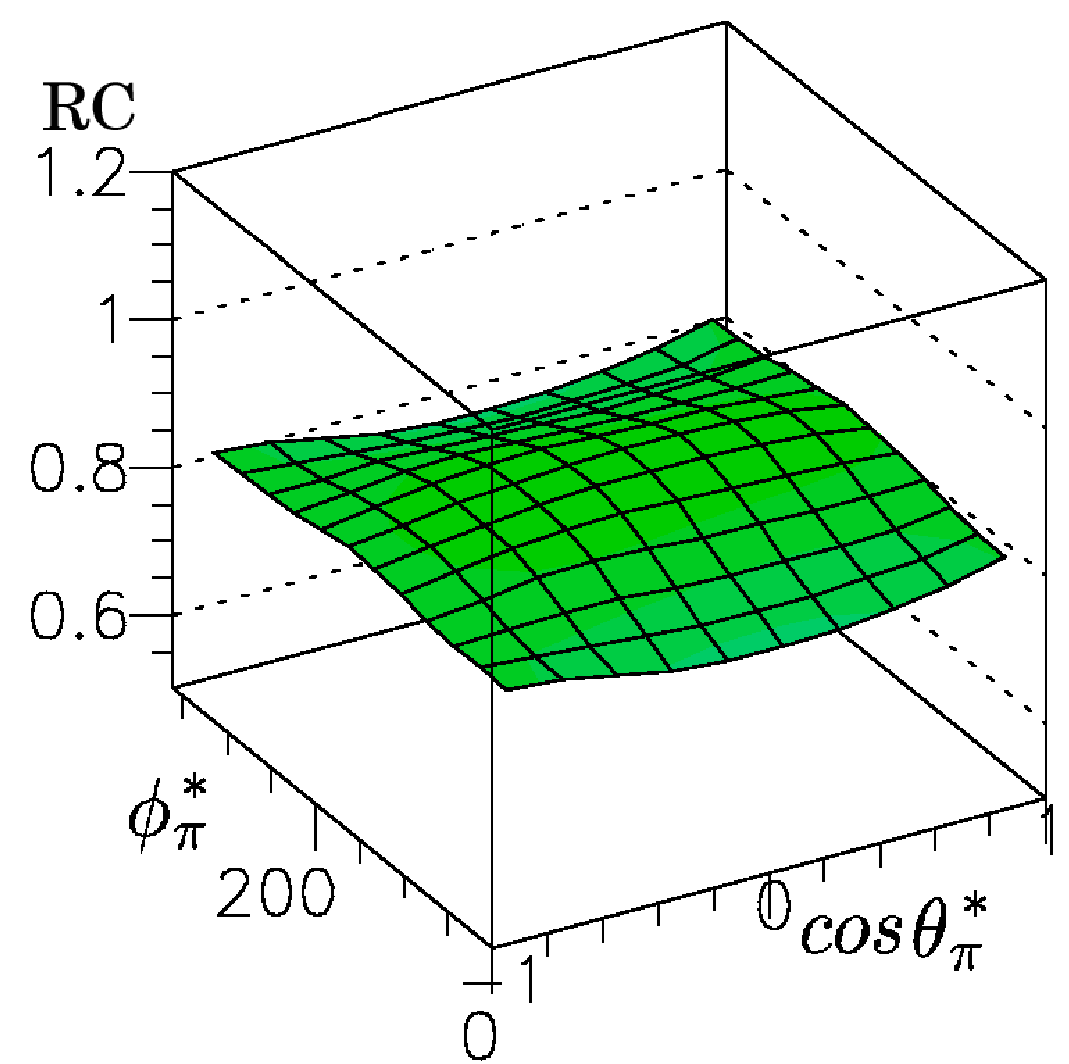}
	\includegraphics[angle=0,width=0.23\textwidth]{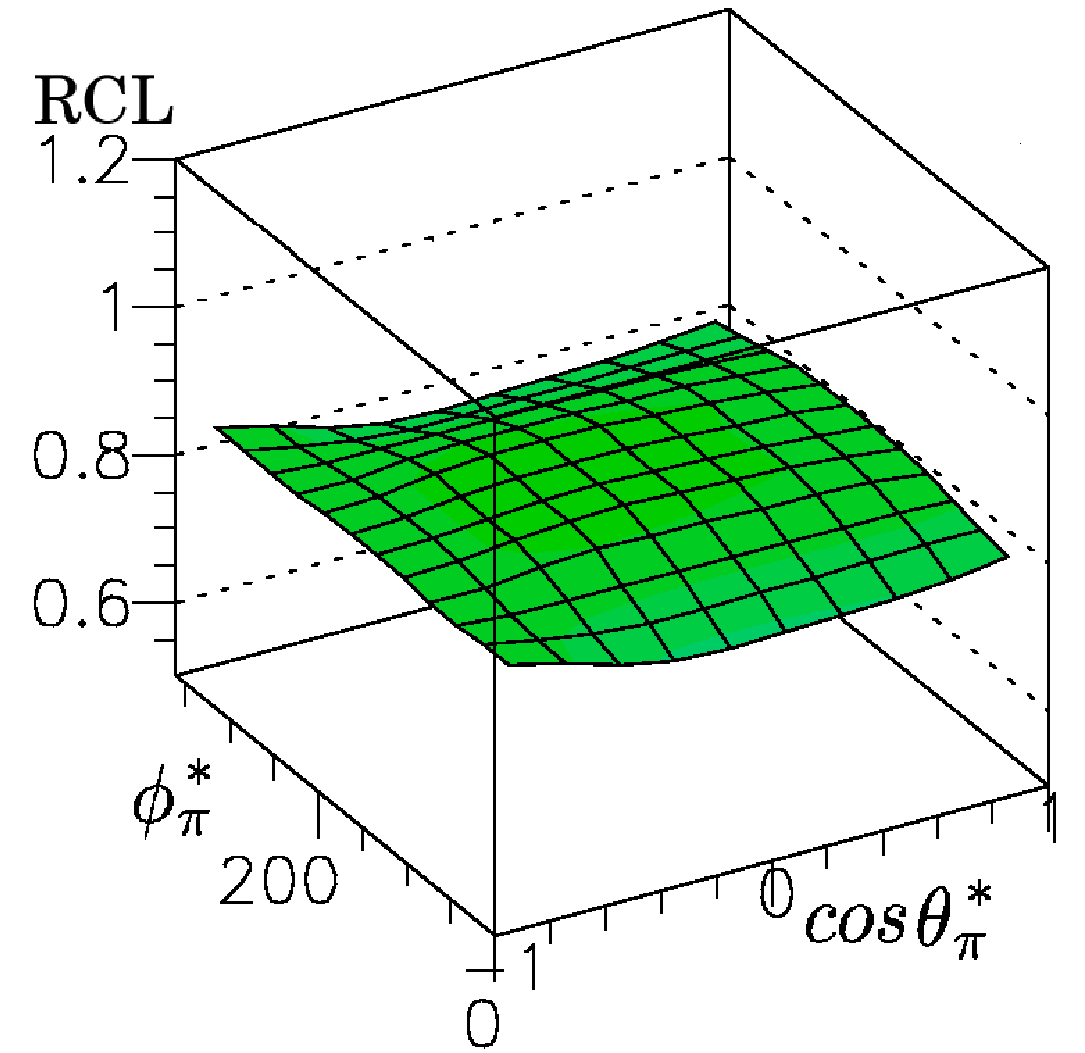}
	\caption{Examples of {\it ExcluRad} results of radiative correction factors for the pion production cross section at a specific kinematics in $W=1.40~\rm{GeV}$, $Q^2=1.7~\rm{GeV^2}$. The left graph shows the exact calculation, the right panel shows the leading log approximation.}
	\label{fig:rad_corr_ratio}
\end{center}
\end{figure}

Radiative processes affect the measured cross section for inclusive electron scattering. They can also modify the measured angular distributions of the hadronic final state. Therefore, a model input that closely reflects the unradiated 5-fold differential hadronic cross section is important. MAID03 was used as model input in a first step, and parameters were adjusted subsequently to optimize the procedure. Figure~\ref{fig:rad_corr_ratio} shows as an example the $\cos\theta^*_{\pi}$ and $\phi^*_{\pi}$ dependences of the correction factor from exact calculation(left) and leading log approximation(right) from ExcluRad. The calcultion agrees with each other within 5\%.

\begin{eqnarray}
RC = \frac{\sigma_{rad}}{\sigma_{norad}}
\end{eqnarray} 
for fixed $W$ and $Q^2$, where $\sigma_{rad}$ is the radiative cross section, and $\sigma_{norad}$ is the un-radiated cross section.  At fixed $Q^2$ and $W$ in the $\Delta(1232)$ region, the radiative corrections are of the order of 20\% and have a visible effect on the angular distribution in the hadronic center-of-mass. 

\subsection{Normalization}

The large acceptance of CLAS and the inclusive electron trigger used in the measurement allowed us to measure elastic and inelastic inclusive electron proton scattering simultaneously with the exclusive process. This allowed us to compare our results with elastic cross sections obtained in dedicated experiments, and to cross check our electron detection efficiency obtained from simulations of the CLAS response. The cross section for the elastic process $ep \to e(p)$ is well known, and parameterizations of its angular dependence can be compared with our measured cross sections. In comparison with the parameterization of Bosted~\cite{Bosted_elastic}, deviations of less than 5\% are observed. We also compared our inclusive inelastic cross section with two parameterizations by Brasse~\cite{FWBrasse} and Keppel~\cite{CKeppel}. The two parameterizations agree well with each other for $W = 1.27 - 1.45~\rm{GeV}$, while there are discrepancies between the two parameterizations at and below the $\Delta$ resonance and in the resonance region at and above 1.5$~\rm{GeV}$. We find excellent agreement at all $Q^2$ with our data in regions where the two parameterizations agree with each other. From the elastic and inelastic cross section measurements, we conclude that the overall normalization uncertainty of this measurement is about $5\%$.

\subsection{Bin centering corrections}

As the cross section can vary significantly within a given kinematics bin, the center of that bin may not coincide with the cross section weighted average within that bin. Corrections are applied using MAID03 as a reasonable representation of these variations. The effects on the cross sections are found to be small, typically much less than $\pm 1.5\%$.

\section{Results and Discussion \label{sec:results}}

\subsection{Differential cross sections}

The five-fold differential cross section for single pion electroproduction is given by following Eq.~\ref{eq:xsec10} :

\begin{eqnarray}\label{eq:xsec10}
&&\frac{\partial^5\sigma}{\partial E_f \partial\Omega_f \partial\Omega_e} = \\
&&\frac{1}{2\pi} \sum \frac{1}{L \;Acc \;\epsilon_{CC}} \frac{1}{\Delta W \;\Delta Q^2 \;\Delta \cos\theta^*_{\pi} \;\Delta \phi^*_{\pi}} \frac{d(W,Q^2)}{d(E_f,\cos\theta_e)} ~,\nonumber
\end{eqnarray}
\noindent 
where  $L$ is the integrated luminosity and $\epsilon_{CC}$ is the efficiency of the $\check{\rm C}$erenkov counter. The last term is the Jacobian:
\begin{equation}\label{eq:jacobian}
\frac{d(W,Q^2)}{d(E_f,\cos\theta_e)} = \frac{2 M_p \;E_i \;E_f}{W}~.
\end{equation}

As mentioned earlier, a 2.5\% minimum acceptance cut was applied to all bins. This cut was used to eliminate bins near the acceptance boundaries and in regions where pion scattering from the torus coils could influence the simulated acceptance.   

Due to the large number of kinematic bins, the resulting differential cross section values cannot be presented in full in this paper. The complete set of cross sections are tabulated in the CLAS Physics Data Base~\cite{clas_db}. In this article we only present examples for the $\phi^*_{\pi}$ and $W$ dependences of the differential cross sections. 
From Eq.~\ref{eqn:cs_form} it is clear that the general structure of the differential cross section for single pion production with unpolarized electrons can be written as:
\begin{eqnarray}
\frac{d\sigma}{d\Omega^*_{\pi}} = A + B\cos{2\phi^*_{\pi}} + C\cos{\phi^*_{\pi}}~.
\label{eq:phi-dependence}
\end{eqnarray}
By fitting the $\phi^*_{\pi}$-dependence of the cross section we can extract the coefficients $A,~B,~C$, which depend on $Q^2$, $W$, and $\cos{\theta^*_{\pi}}$ only. They are related to the various cross section pieces as given in the following equations:
\begin{eqnarray}
 A &=& \sigma_T + \epsilon \sigma_L\\
 B &=& \epsilon \sigma_{TT} \\
 C &=& \sqrt{2\epsilon(1+\epsilon)} \sigma_{LT}~.
\end{eqnarray}      

\begin{figure}[!thb]
\begin{center}
	\includegraphics[angle=0,width=0.48\textwidth]{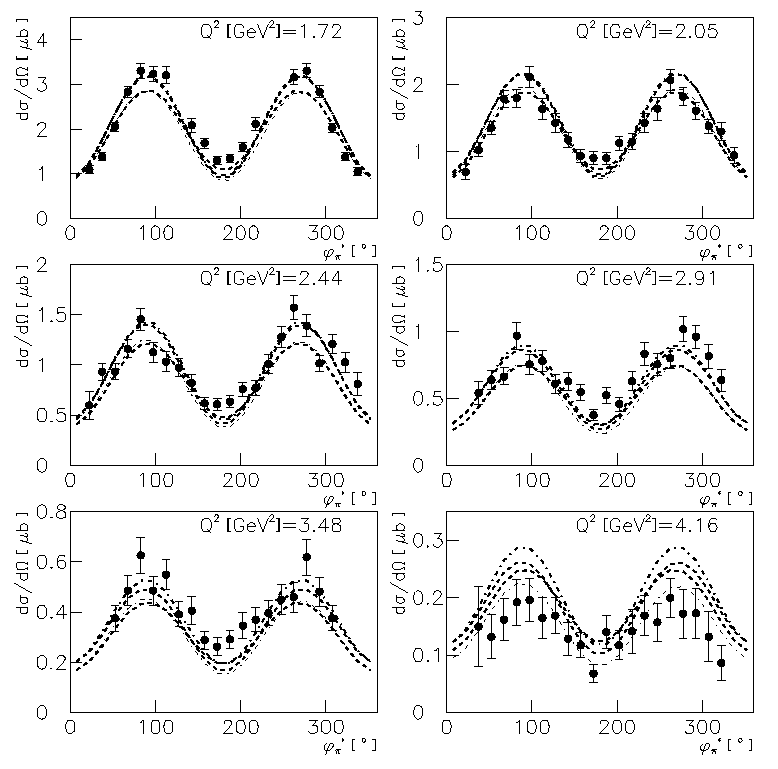}
	\caption{
          Differential cross section vs $\phi^*_{\pi}$  in the $\Delta(1232)$  region at fixed $\cos \theta^*_{\pi} =-0.1$ for different bins in $Q^2$. DMT(bold dash), MAID00(thin dash), MAID03(bold dash-dot) and SL04(thin dash-dot). The error bar of data shows only statistical error.
	\label{fig:cs1}}
\end{center}
\end{figure}

\begin{figure}[!thb]
\begin{center}
	\includegraphics[angle=0,width=0.48\textwidth]{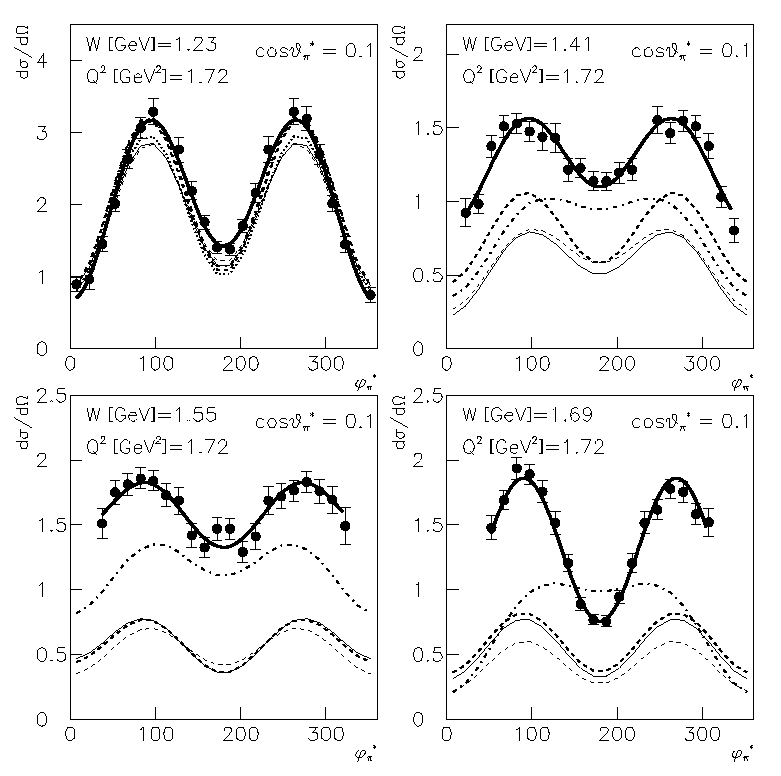}
	\caption{
          Examples of differential cross section [$\mu b/sr$] at fixed $Q^2$, $\cos\theta^*_{\pi}$ vs $\phi^*_{\pi}$ for different $W$. DMT(bold dash), MAID98(thin solid), MAID00(thin dash), MAID03(bold dash-dot), SL(bold dot )and SL04(thin dash-dot). The bold solid line is the result of the $A + B \cos \phi^*_{\pi}  + C \cos 2\phi^*_{\pi} $ fit to the data .
	\label{fig:cs1-1}}
\end{center}
\end{figure}


\begin{figure}[!thb]
\begin{center}
	\includegraphics[angle=0,width=0.48\textwidth]{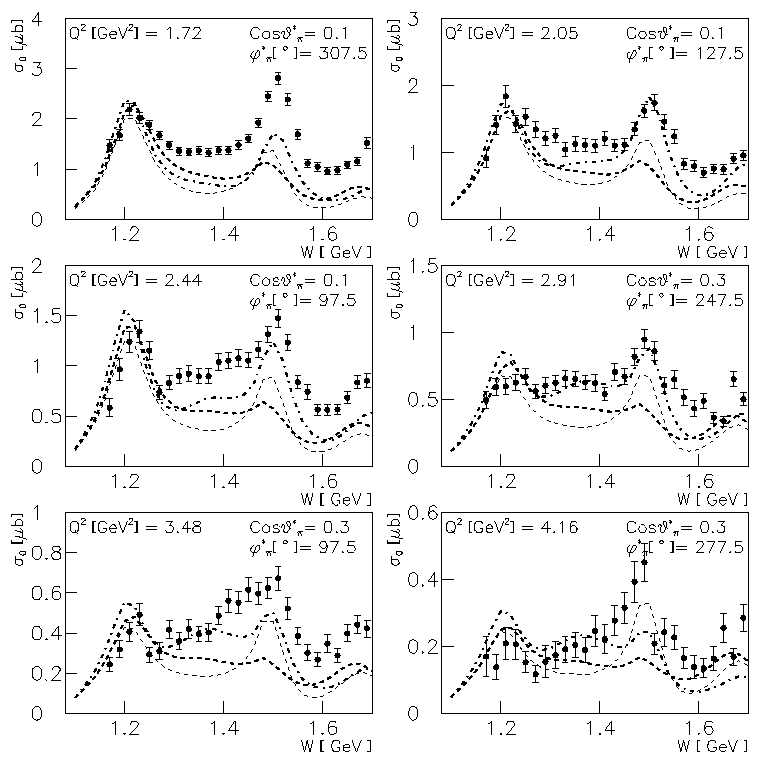}
	\caption{
         Samples of differential cross section vs $W$ at different $\cos\theta^*_{\pi}$ and $\phi^*_{\pi}$. Curves as in Fig.~\ref{fig:cs1}. $\sigma_0$ is definded by Eq. \ref{eq:phi-dependence}.
	\label{fig:cs2}}
\end{center}
\end{figure}


In Fig.~\ref{fig:cs1} and Fig.~\ref{fig:cs1-1} the $\phi^*_{\pi}$-dependence of the differential cross section is shown for various $Q^2$ and $W$ values at fixed $\cos{\theta^*_{\pi}}$, and compared with models discussed in Section~\ref{sect:models}.
In the $\Delta(1232)$ region and at the lower $Q^2$, the models are close to each other, and in general, give a good description of the shape of the data. However, the relatively good model description of the $\Delta(1232)$ region is largely due to the fact that the resonance contributions are known from the analysis of $p\pi^0$ electroproduction in that mass range and have been incorporated into the models. The $p\pi^{\circ}$ channel is more sensitive to isospin $\frac{3}{2}$ resonances, and it also has less strength in the nonresonant amplitudes. At high $Q^2$ there are discrepancies with the models discussed for near in-plane azimuthal angles, indicating that nonresonant contributions may not be fully represented in the model calculations. The discrepancies between the models and the data become larger with increasing $W$, clearly showing that there is significant strength missing in the mass region above the $\Delta(1232)$. This indicates that the strengths of some of the higher mass resonances are underestimated in the models. In Fig.~\ref{fig:cs1-1} cross sections are displayed at fixed $Q^2$ and $\cos\theta^*_{\pi}$ for different regions in $W$ where resonance contributions should be maximum.  The $W$ dependence of the differential cross section is shown in Fig.~\ref{fig:cs2} for selected $\phi^*_{\pi}$ and $\cos\theta^*_{\pi}$ bins and different $Q^2$ values. The $\Delta(1232)$ and the resonances around $W=1.5~\rm{GeV}$ are clearly seen at all $Q^2$.    

\subsection{Electron Beam Asymmetry} 
The experiment was performed with a highly longitudinally polarized electron beam on an unpolarized hydrogen target. The beam polarization allows access to the structure function $\sigma_{LT'}$, which can be separately determined by a measurement of the polarized beam asymmetry for the yield of pions produced by electrons with helicity aligned parallel and anti-parallel to the beam direction. The asymmetry can be written as 
\begin{equation}\label{eqn:altp2}
A_{LT'} = \frac{N^+_{\pi} - N^-_{\pi}}{P_e (N^+_{\pi}+ N^-_{\pi})}~,
\end{equation}
where $P_e$ is the electron polarization and $N_{\pi}^{\pm}$ is the measured number of pion events in a specific kinematic bin for $\pm$ electron beam helicity states after applying all corrections. In order to obtain the $N_{\pi}^{\pm}$ for each bin, corrections for the beam charge asymmetry, radiative effects, and binning effects have been applied. The beam charge asymmetry(BCA) may result from a helicity-dependent current variation present in the beam. The BCA was measured and monitored continuously using the charge information from the Faraday cup and other beam monitors. In addition, the inclusive elastic and inelastic electron scattering rates were measured continuously. These rates were normalized to the integrated charge for each helicity. The only physics process that can produce an inclusive asymmetry is parity violation, which is several orders of magnitudes smaller than the asymmetries observed in exclusive pion production and can be neglected. 
Radiative corrections(RC) were applied by calculating RC$^{\pm}$ for each helicity state. This was accomplished using MAID00 as the model cross section input.  MAID03 was used to study systematic uncertainties, which were found to be less than 1\%. The corrected asymmetry is given by 
\begin{eqnarray}
A_{LT'} &=& \frac{N^+_{\pi}/RC^+ - N^-_{\pi}/RC^-}{N^+_{\pi}/RC^+ + N^-_{\pi}/RC^-} \\
RC^{\pm} &=& \frac{\sigma^{\pm}_{rad}}{\sigma^{\pm}_{norad}}~,
\end{eqnarray}
where $\sigma^{\pm}_{rad}$ and $\sigma^{\pm}_{norad}$ are the radiated and nonradiated model cross sections for each helicity state. Bin centering corrections were found to be negligible in the asymmetry.

\begin{figure}[!thb]
\begin{center}
	\includegraphics[angle=0,width=0.5\textwidth]{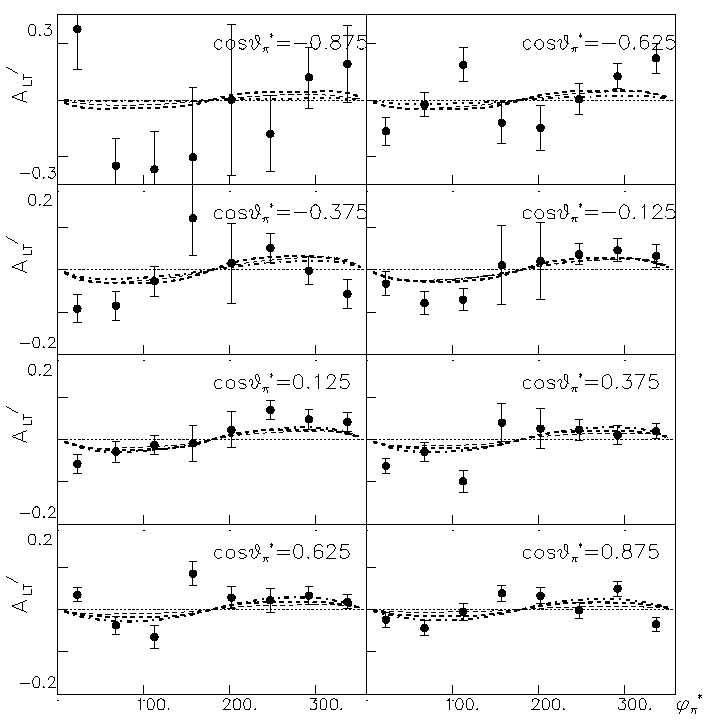}
	\includegraphics[angle=0,width=0.5\textwidth]{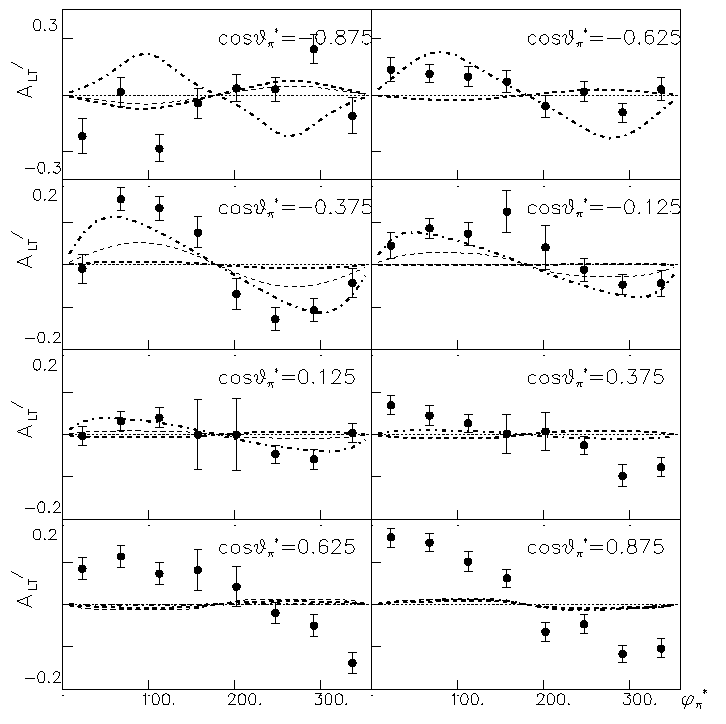}
	\caption{\small 
         Electron beam helicity asymmetry as a function $\phi^*_{\pi}$ in different $\cos\theta^*_{\pi}$ bins at $W=1.40$~$\rm{GeV}$(top) and $W=1.69$~$\rm{GeV}$(bottom), $Q^2=2.05$~$\rm{GeV^2}$ compared to different physics models  MAID98(thin solid), MAID00(thin dash), MAID03(bold dash-dot), and DMT(bold dash)
	\label{fig:asym_f15}}
\end{center}
\end{figure}

Examples of the electron asymmetry $A_{LT'}$ are shown in Fig.~\ref{fig:asym_f15}. We can see the sensitivity of this quantity to different models. None of the models gives a satisfactory description of $A_{LT'}$ for all angle bins. At the higher $W$ the comparison shows a strong model sensitivity. Large discrepancies are seen at forward angles and high $W$. This could be due to an underestimation of t-channel processes.

\subsection{Systematic uncertainties}

The main contributions to the systematic uncertainties of this measurement are due to the cuts applied to identify electrons and positively charged pions and the definition of the fiducial volumes. They were studied by individually changing the cut values for the electron and pion selection to provide more stringent or less stringent particle selections, and redoing the complete analysis. This resulted in global estimates of the systematic uncertainties of $\pm 5.9$\%.

 All other contributions are at the 1\% level or below. These include uncertainties due to the missing mass cut( cut at $\pm 3\sigma$ vs cut at $\pm 2 \sigma$ from peak maximum), uncertainties in the target length and the liquid-hydrogen density, and sensitivity of the acceptance calculations to the specific model(MAID00 vs MAID03) used in the simulation. In addition, the effect of radiative correction on the cross section was studied using different model parameterizations(MAID03 vs MAID00) for the corrections.  Adding the systematic uncertainties in quadrature results in a global systematic uncertainty of 6.3\%. However, in order to study the systematic uncertainties for all bins in cross section and asymmetry, the complete cross section extraction and asymmetry analyses were repeated and systematic uncertainties determined for every data point. 

\subsection{Structure Functions}
\begin{figure}[!thb]
\begin{center}
	\includegraphics[angle=0,width=0.49\textwidth]{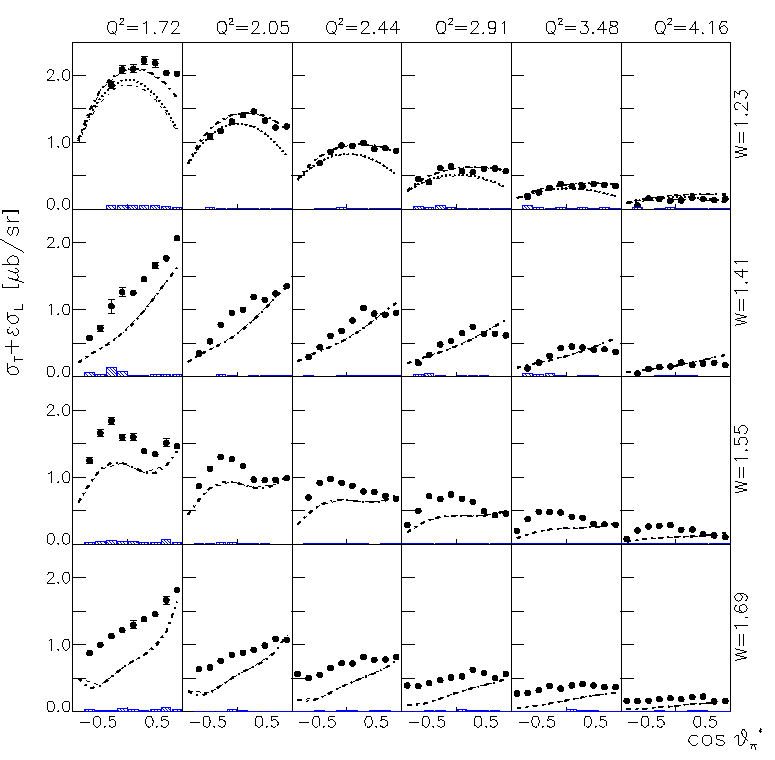}
	\caption{\small Structure function $\sigma_T + \epsilon \sigma_L$ as a function of $\rm{cos}\;{\theta^*_{\pi}}$ for different  $W$ and $Q^2$ values in comparison with  model calculations  MAID00(thin dash), MAID03(bold dash-dot), SL(bold dot) and SL04(thin dash-dot). The shaded bars shows the systematic uncertainties.
	\label{fig:sigma_l_t01}}
\end{center}
\end{figure}

The unpolarized differential cross section contains four structure functions. By fitting the $\phi^*_{\pi}$-dependence we can extract  $\sigma_T + \epsilon\sigma_L$, and the two structure functions $\sigma_{TT}$, $\sigma_{LT}$. The structure function $\sigma_{LT'}$ is determined by fitting $\sigma_0\cdot A_{TL^{\prime}}$ with the form $a \sin\phi^*_{\pi}$, where $a = \sqrt{2\epsilon (1-\epsilon)}\sigma_{LT'}$.          
Figure~\ref{fig:sigma_l_t01} shows the combination of structure functions $\sigma_T + \epsilon \sigma_L$ versus $\cos{\theta^*_{\pi}}$ in four $W$ bins near the masses of four prominent resonances, the $\Delta(1232)$, the Roper resonance $N(1440)$, the $N(1535)$ and $N(1680)$, and for different $Q^2$ values. The numerical results of this fit for the total cross section $\sigma_T + \epsilon \sigma_L$ are tabulated in the Appendix. 

We see that at $W$ values above the $\Delta(1232)$, the models underestimate the total virtual photon absorption cross section. For most models, this is even the case in the $\Delta(1232)$ region. In the higher mass regions, both versions of MAID underpredict the global strength significantly. With the exception of the very forward region, where there is good agreement with the data, the models account for only about 50\% of the strength at larger angles. Since there is little variation from model to model, it is difficult to discuss the origin of the discrepancy with the data. However, since the structures in the angular distribution at $W = 1.55$~$\rm{GeV}$ are not well reproduced, most likely resonance contributions in the second resonance region are underestimated in the models. 

\begin{figure}[!thb]
\begin{center}
	\includegraphics[angle=0,width=0.49\textwidth]{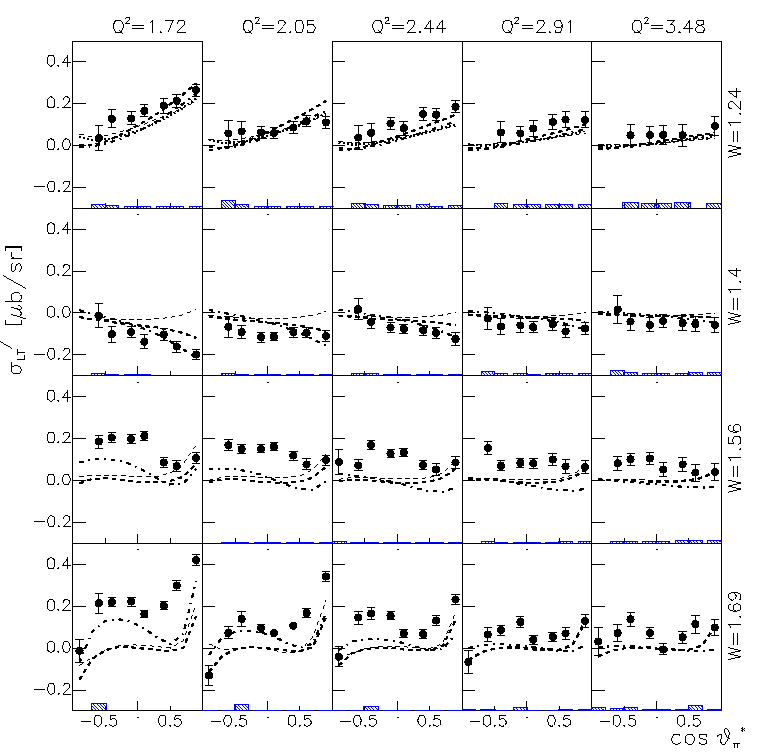}
	\caption{\small Structure function $\sigma_{LT'}$ as a function of $\rm{cos}\;{\theta^*_{\pi}}$ at fixed $W$ and $Q^2$ compared to model calculations. Curves as in Fig.~\ref{fig:sigma_l_t01}. Systematic uncertainties are shown in shaded bars.
	\label{fig:sigmaltp01}}
\end{center}
\end{figure}

The polarized beam structure function $\sigma_{LT'}$ in  Figure~\ref{fig:sigmaltp01} exhibits more sensitivity to models. In the $\Delta(1232)$ region, $\sigma_{LT'}$ is positive and increases towards forward angles. The model dependence is small, and a reasonable description of the angular dependence is provided by all models. In the region of the Roper resonance, $\sigma_{LT'}$ is negative and rises in magnitude for forward angles. The fourth row shows $\sigma_{LT'}$ in the 3rd resonance region, which is dominated by the $F_{15}(1680)$. Here  $\sigma_{LT'}$ changes sign again, and the angular dependence shows more structure. There is a significant backward enhancement and a forward peak at lower $Q^2$.  The former indicates the presence of strong resonance contributions, while the latter shows the importance of t-channel processes contributing to the background amplitudes. While MAID03 shows qualitatively a similar behavior, the structures are not quantitatively reproduced in any model.

\begin{figure}[!thb]
\begin{center}
	\includegraphics[angle=0,width=0.49\textwidth]{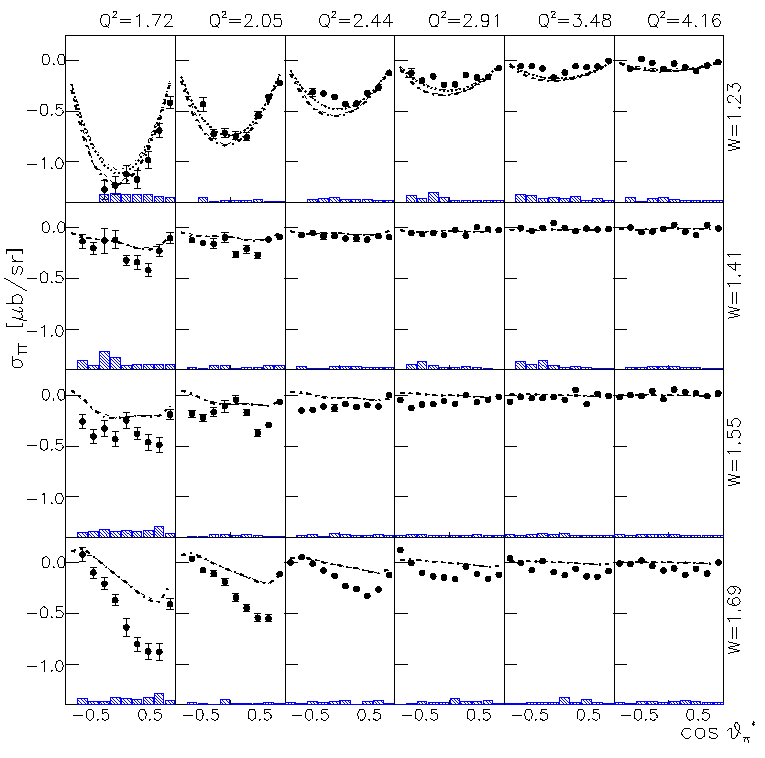}
	\caption{\small Structure function $\sigma_{TT}$ as a function of $\rm{cos}\;{\theta^*_{\pi}}$ at fixed $W$ and $Q^2$ compared to model calculations. Curves as in Fig.~\ref{fig:sigma_l_t01}. Systematic uncertainties are shown in shaded bars.
	\label{fig:sigma_tt_01}}
\end{center}
\end{figure}
\begin{figure}[!thb]
\begin{center}
	\includegraphics[angle=0,width=0.49\textwidth]{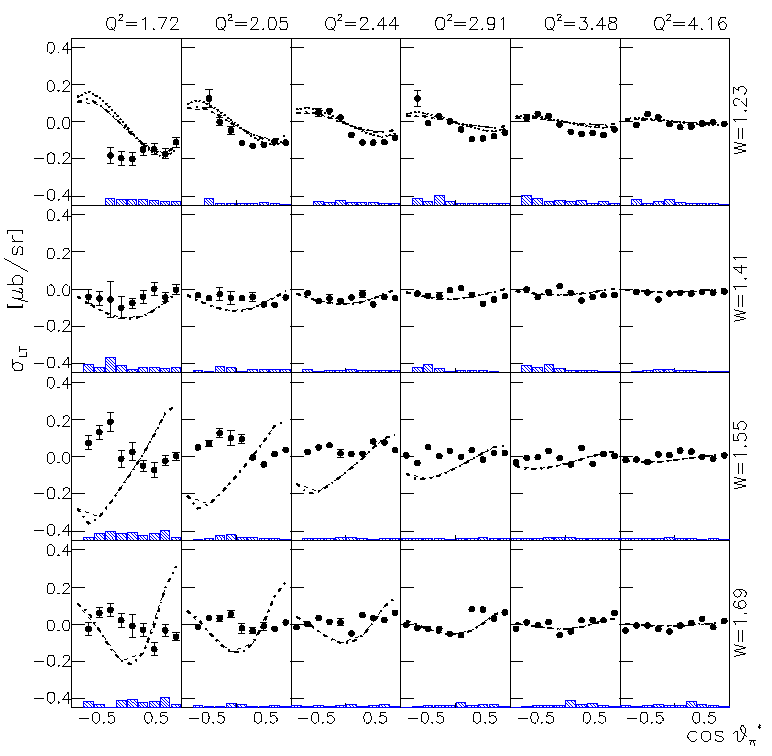}
	\caption{\small Structure function $\sigma_{LT}$ as a function of $\rm{cos}\;{\theta^*_{\pi}}$ at fixed $W$ and $Q^2$ compared to model calculations. Curves as in Fig.~\ref{fig:sigma_l_t01}. Systematic uncertainties are shown in shaded bars.
	\label{fig:sigma_lt_01}}
\end{center}
\end{figure}

Some of the $\sigma_{TT}$ and the $\sigma_{LT}$  interference structure functions are shown in Fig.~\ref{fig:sigma_tt_01} and in Fig.~\ref{fig:sigma_lt_01} for various $W$ and $Q^2$ values. For $\sigma_{TT}$, the models show qualitatively a similar behavior but underestimate the magnitude. For $\sigma_{LT}$ there is qualitative disagreement between models and the data at nearly all kinematics. At the larger $W$ even the signs are different. 

\subsection{Moments of Legendre polynomials}
The full impact of the data presented in this paper on the extraction of the nucleon resonance transition form factors can only be obtained in a global partial wave analysis that also incorporates data on other reaction channels such as $e p \to e p \pi^{\circ}$ and $e p \to e p \eta$. It also requires theoretical input on the contributing background amplitudes. This work is the subject of a forthcoming paper~\cite{aznauryan_2007}. However, some insight into the dominant partial waves contributing to the reaction can be obtained from a Legendre polynomial expansions of the structure functions. The structure functions can be formally written as sums of Legendre polynomials. 
\begin{eqnarray}
\sigma_T+ \epsilon \sigma_L = \sum_{l=0}^{n} D_l^{T+L}P_l(\cos{\theta^*_{\pi}})\\
\sigma_{LT} = \sin{\theta^*_{\pi}} \sum_{l=0}^{n-1} D_l^{LT} P_l(\cos{\theta^*_{\pi}})\\
\sigma_{LT'} = \sin{\theta^*_{\pi}} \sum_{l=0}^{n-1} D_l^{LT'} P_l(\cos{\theta^*_{\pi}})\\
\sigma_{TT} = \sin^2{\theta^*_{\pi}} \sum_{l=0}^{n-2} D_l^{TT} P_l(\cos{\theta^*_{\pi}})~,
\end{eqnarray}     
where $P_l(\cos{\theta^*_{\pi}})$ is the $l^{th}$-order Legendre polynomial, and the $D_l$'s are the Legendre moments. For single pion electroproduction, each moment can be written as an expansion in magnetic($M_{l_{\pi^{\pm}}}$),   electric($E_{l_{\pi^{\pm}}}$), and scalar($S_{l_{\pi^{\pm}}}$) multipoles~\cite{multipole00}. A complete global analysis will have to include all relevant multipoles. However, when going to sufficiently high $l_{\pi}$, these expressions become rather unwieldy and are not discussed here.      

\begin{figure}[!thb]
\begin{center}
	\includegraphics[angle=0,width=0.48\textwidth]{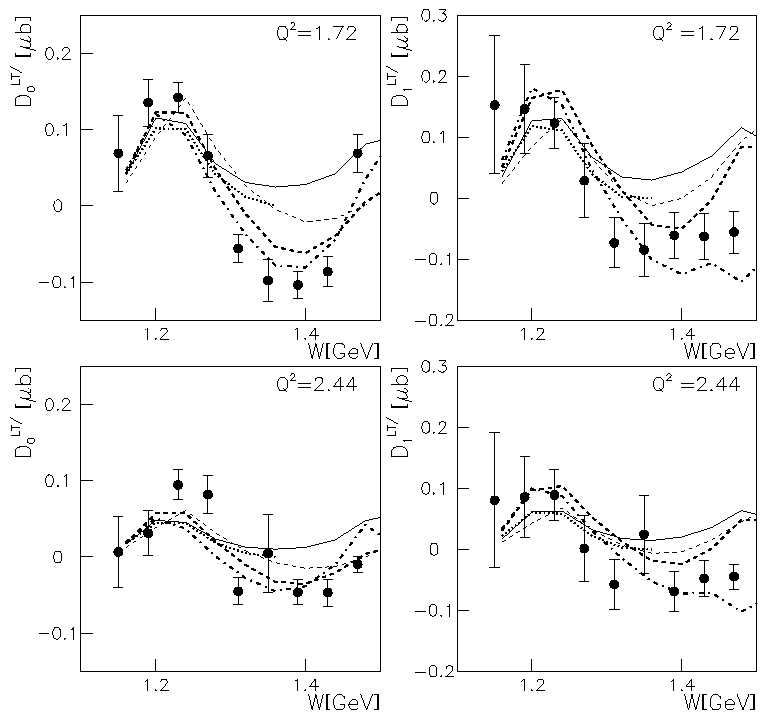}
	\caption{\small Legendre moment $D_0^{LT'}$, $D_1^{LT'}$ of $\sigma_{LT'}$ vs $W$. Model predictions: MAID98(thin solid), MAID00(thin dash), MAID03(bold dash-dot), SL(bold dot), and DMT(bold dash)
	\label{fig:lmoment_w1}}
\end{center}
\end{figure}
\begin{figure}[!thb]
\begin{center}
	\includegraphics[angle=0,width=0.49\textwidth]{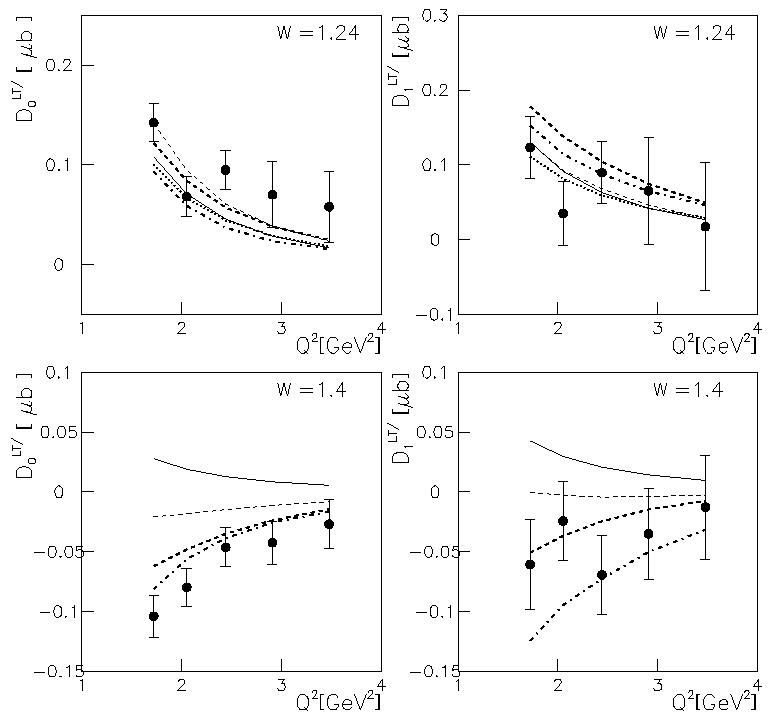}
	\caption{\small The $Q^2$ dependence of Legendre moments of $\sigma_{LT'}$ for $\pi^+$ channel at $W=1.24,\;1.4\;\rm{GeV}$. Curves as in Fig.~\ref{fig:lmoment_w1}.
	\label{fig:lmoment}}
\end{center}
\end{figure}

Figure~\ref{fig:lmoment_w1} shows the $W$-dependence of the fitted $D_0^{LT'}$ and $D_1^{LT'}$ Legendre moments. Both moments follow the strong resonant behavior in the $\Delta(1232)$ region, and change sign between the $\Delta(1232)$ and the second resonance region. The comparison with the MAID03 and DMT models strongly hints that the sign change and strong negative amplitude is due to the significantly increased strength of the Roper resonance compared with the earlier versions of MAID. This indicates a strong sensitivity of the polarized structure function $\sigma_{LT'}$ to the interference of the Roper multipoles with background amplitudes.  
In Fig.~\ref{fig:lmoment} we show the $Q^2$-dependence of the moments for two $W$ values near the $\Delta(1232)$ and the Roper $N(1440)$ resonances. The strongest model dependence at the $\Delta(1232)$ mass is seen in the $D_1^{LT'}$ moment, where the data have a slight preference for the SL model and the previous MAID versions over  MAID03 and DMT. Near the mass of the Roper resonance all moments show a strong model dependence which, as mentioned earlier, is largely due to the different strength in the amplitudes of the Roper resonance. MAID03 and DMT give a good description for $D_0^{LT'}$ and $D_1^{LT'}$, while the older MAID versions fail to fit the data.    
\begin{figure}
	\includegraphics[angle=0,width=0.49\textwidth]{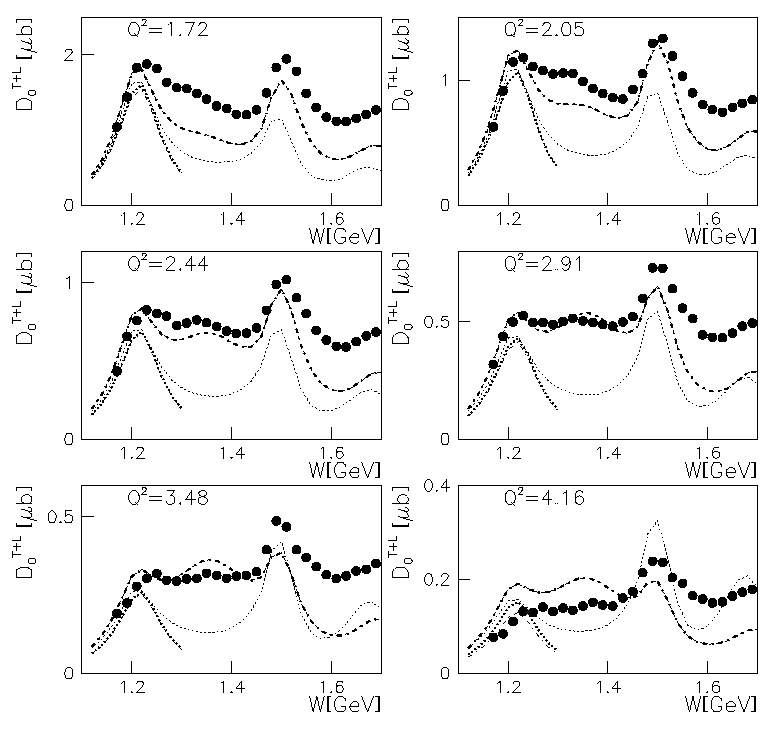}
	\caption{\small Legendre moment $D_0^{T+L}$ of structure function $\sigma_T + \epsilon \sigma_L$ vs $W$. Curves: MAID03(bold dash-dot), MAID00(thin dash), SL(bold dot), SL04(thin dash-dot) and MAID00 with  the $P_{11}(1440)$ amplitudes switched off(thin dot).
	\label{fig:lmoment_0_sigma0}}
\end{figure}
\begin{figure}[!thb]
\begin{center}
	\includegraphics[angle=0,width=0.49\textwidth]{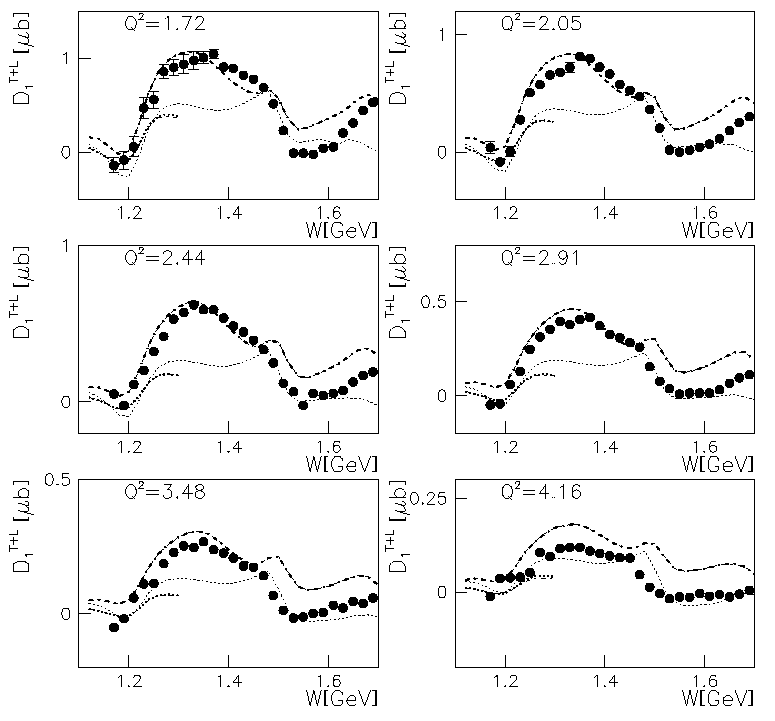}
	\caption{\small Legendre moment $D_1^{T+L}$. Curves as in Fig.~\ref{fig:lmoment_0_sigma0}.
	\label{fig:lmoment_1_sigma0}}
\end{center}
\end{figure}
\begin{figure}[!thb]
\begin{center}
	\includegraphics[angle=0,width=0.49\textwidth]{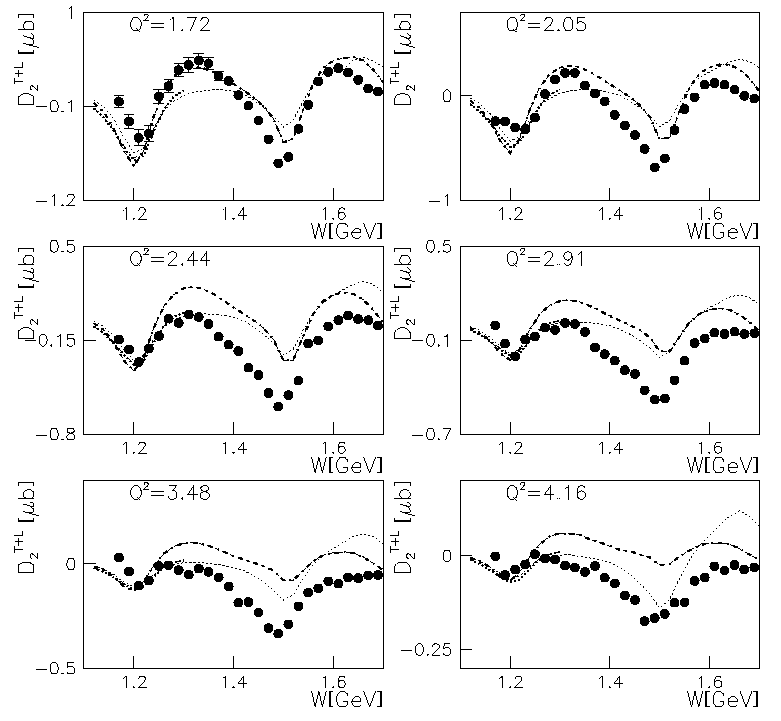}
	\caption{\small Legendre moment $D_2^{T+L}$ vs $W$. Curves as in Fig.~\ref{fig:lmoment_0_sigma0}. 
	\label{fig:lmoment_2_sigma0}}
\end{center}
\end{figure}

Figures ~\ref{fig:lmoment_0_sigma0},~\ref{fig:lmoment_1_sigma0}, and~\ref{fig:lmoment_2_sigma0} show the three lowest order Legendre moments of the structure function $\sigma_T + \epsilon \sigma_L$ vs $W$ and for different $Q^2$. $D_0^{T+L}$ projects out the $\cos{\theta^*_{\pi}}$-independent part of the contributing partial waves. Resonance structure is clearly visible in the $\Delta(1232)$ region and near 1.5 $\rm{GeV}$. 
The enhancement near 1.5 $\rm{GeV}$ is obviously related to the $D_{13}(1520)$ and $S_{11}(1535)$ states, while the increase in strength near 1.7 $\rm{GeV}$ structure is related to several states with the $F_{15}(1680)$ being a dominant contribution. At higher $Q^2$, the resonant structure near 1.5 $\rm{GeV}$ becomes increasingly dominant in comparison with the $\Delta(1232)$. The broad shoulder between the $\Delta$ and the 1.5 $\rm{GeV}$ peak is related to the Roper resonance $P_{11}(1440)$, which also becomes more prominent with increasing $Q^2$. This is clearly seen when the Roper amplitudes are switched off in the models. MAID gives a qualitative description of this region but underestimates the magnitude at the lower $Q^2$. The $\Delta(1232)$ and the $D_{13}(1520)$ resonances show most clearly in the $D_2^{T+L}$ Legendre moment. Also, the increasing prominence of the $D_{13}(1520)$ over the $\Delta(1232)$ at high $Q^2$ is clearly visible. The $D_1^{T+L}$ moment is dominated by the Roper $J^P = \frac{1}{2}^+$ amplitudes that interfere with background amplitudes resulting in the broad structure extending from 1.25 to 1.5 $\rm{GeV}$. This structure is quite well described by MAID. However, MAID overestimates the strength in this moment at higher $Q^2$.     

\begin{figure}[!thb]
\begin{center}
	\includegraphics[angle=0,width=0.48\textwidth]{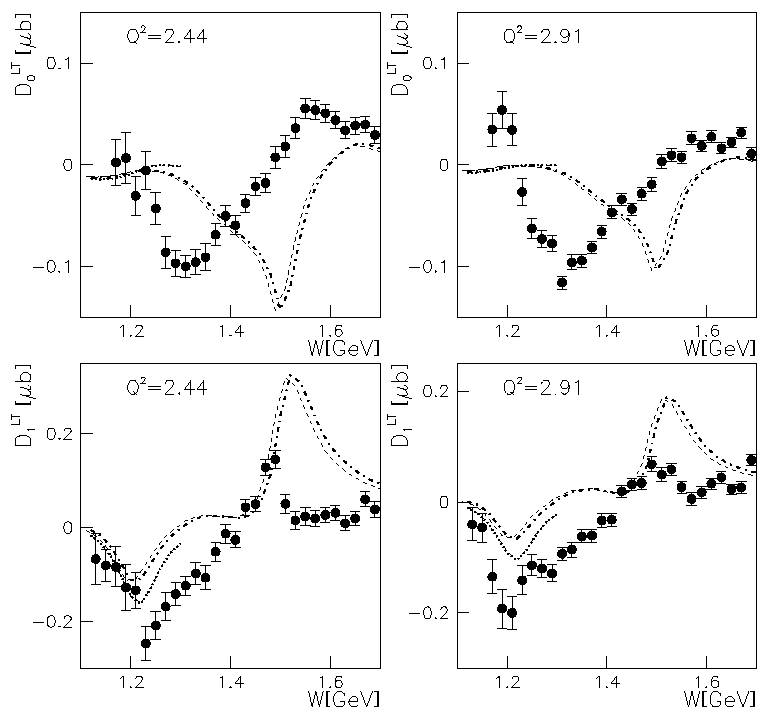}
	\caption{\small Legendre moments $D_0^{LT}$(top) and $D_1^{LT}$(bottom) of structure function $\sigma_{LT}$. Curves as in Fig.~\ref{fig:lmoment_0_sigma0}.
	\label{fig:lmoment_sigma_LT}}
\end{center}
\end{figure}
\begin{figure}[!thb]
\begin{center}
	\includegraphics[angle=0,width=0.49\textwidth]{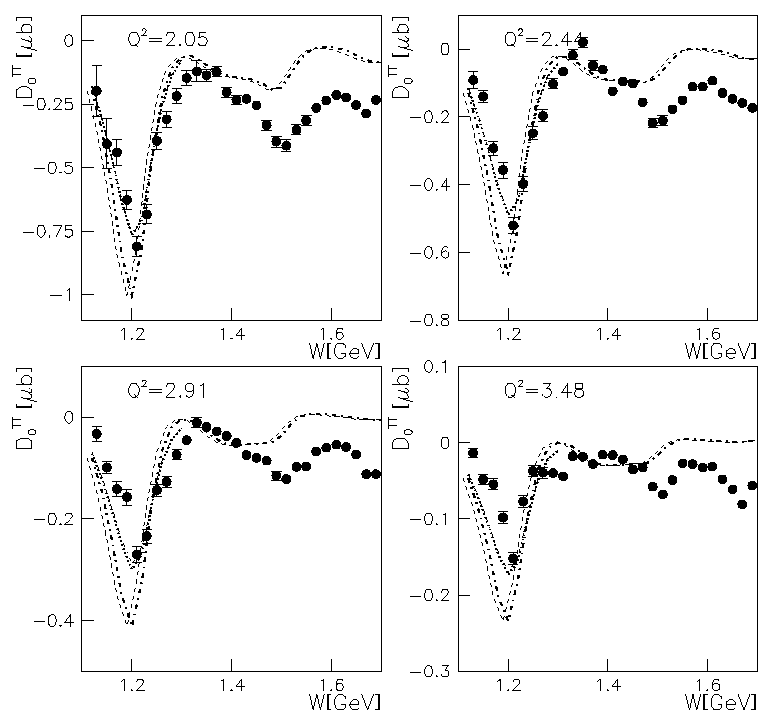}
	\caption{\small Legendre moment $D_0^{TT}$ of the structure function $\sigma_{TT}$ vs $W$. Curves as in Fig.~\ref{fig:lmoment_0_sigma0}.
	\label{fig:lmoment_sigma_TT}}
\end{center}
\end{figure}

The lowest-order moments of $\sigma_{LT}$ are shown in Fig.~\ref{fig:lmoment_sigma_LT}. Both moments show a zero-crossing near 1.45 $\rm{GeV}$. MAID predicts a sign change only for $D_1$. For both $D_0^{LT}$ and $D_1^{LT}$, MAID predicts resonance-like behavior near 1.5~$\rm{GeV}$ at all $Q^2$, which is absent from the data.  
Figure~\ref{fig:lmoment_sigma_TT} shows the lowest order moment of the transverse interference structure function $\sigma_{TT}$. $D_0^{TT}$ is dominated by the $\Delta(1232)$ structure and also exhibits resonance structure near 1.5 $\rm{GeV}$. The $\Delta(1232)$ is described well by both the SL model and MAID00, while in the 1.5 $\rm{GeV}$ region, MAID shows less resonance strength than the data.

\section{Summary}
In this paper we presented the first high $Q^2$ measurements and complete angular distributions for exclusive $\pi^+$ electroproduction on protons in the nucleon resonance region. The 5-fold differential cross section $\frac{\partial^5\sigma}{\partial E^{\prime} \partial\Omega_e \partial\Omega_{\pi}}$ was measured for 31,295 kinematic bins in a large range of $Q^2$,  $W$, azimuthal angle $\phi^*_{\pi}$, and polar angle $\theta^*_{\pi}$.  In addition, the electron beam asymmetry was measured in the same kinematical range. The differential cross sections and the extracted structure functions show strong sensitivity to model descriptions of the reaction process and reveal significant lack of resonance strength above the $\Delta(1232)$ in all models. The polarized interference structure function $\sigma_{LT'}$ exhibits strong sensitivity to the Roper multipoles interfering with background multipoles. A study of the $W$-dependence of the two lowest Legendre moments for $\sigma_{LT'}$ supports this observation. Many features of the data are described qualitatively by available model parameterizations but lack a quantitative explanation. This is not surprising as all models have been tuned only on single $\pi^0$ production. A striking discrepancy between data and models is seen in the Legendre moments of the $\sigma_{LT}$ structure function, which allows a qualitatively very different behavior from what the models predict. This indicates that some important process is not correctly implemented in the model descriptions. We hope that the data presented here will help remedy the situation.    

The full data set, only a fraction of which was presented here, will serve as input in forthcoming global analyses to extract the $Q^2$-dependence of the transition form factors for several resonances with masses in the range up to 1.7 $\rm{GeV}$. The analysis of our exclusive $\pi^+$ data in a global fit that also includes the single $\pi^0$ and $\eta$ channels, will allow one to separate the isospin $\frac{1}{2}$ and isospin $\frac{3}{2}$ states. These data may also be used to vastly improve the description of resonance production processes and the transition form factors in dynamical models. The complete set of differential cross sections and beam spin asymmeteries are available from the CLAS Physics Data Base~\cite{clas_db}.

We acknowledge the outstanding effort of the staff of the Accelerator and Physics Divisions at Jefferson Lab in their support of this experiment. This work was supported in part by the U.S. Department of Energy and the National Science Foundation,  the Korea Research Foundation, the French Commissariat $\grave{\rm{a}}$ l'Energie Atomique, and the Italian Istituto Nazionale di Fisica Nucleare. The Southeastern Universities Research Association(SURA) operated the Thomas Jefferson National Accelerator Facility for the United States Department of Energy under contract DE-AC05-06OR23177. \\

\newpage

\end{document}